\documentclass[amsmath,amssymb,11pt]{article}
\pdfoutput=1

\usepackage{jheppub}
\usepackage[utf8]{inputenc}
\usepackage{amsfonts}
\usepackage{graphicx}
\usepackage{slashed}
\usepackage{subfig}
\usepackage{array}
\usepackage{hyperref}
\usepackage{braket}
\usepackage{color}
\newcommand{\beq}{\begin{equation}}
\newcommand{\eeq}{\end{equation}}

\title{Novel Aspects of the Extended First Law of Entanglement}

\author[a]{Felipe Rosso}
\author[b]{and Andrew Svesko}

\affiliation[a]{Department of Physics and Astronomy, University of
Southern California, Los Angeles,\\ CA 90089-0484, USA.}
\affiliation[b]{Department of Physics, Arizona State University, 
Tempe, Arizona 85287, USA.}

\emailAdd{felipero@usc.edu}
\emailAdd{asvesko@asu.edu}

\abstract{Using AdS/CFT an extended first law of entanglement has been previously derived for the vacuum reduced to a ball in Minkowski. The statement not only includes perturbations of the state but also of the conformal field theory (CFT), via variations of the generalized central charge. We clarify some subtleties previously overlooked and use simple arguments to generalize prior derivations to arbitrary gravity theories in the bulk as well as new regions in the boundary CFT. Our construction also applies to two-dimensional bulk theories and admits an interesting extension for a three-dimensional bulk, providing a curious result regarding the thermodynamic volume in extended black hole thermodynamics. We discuss future prospects regarding the extended first law of entanglement.}

\begin{document} 

\setcounter{tocdepth}{2}
\maketitle
\flushbottom

%%%%%%%%%%%%%%%%%%%%%%%%%%%%%%%%%%%%%%%%%%%%%
\section{Introduction}
%%%%%%%%%%%%%%%%%%%%%%%%%%%%%%%%%%%%%%%%%%%%%

The first law of entanglement is a natural generalization of the first law of thermodynamics that applies to non-equilibrium states. As first shown in Refs. \cite{Blanco:2013joa,Wong:2013gua}, it is a consequence of positivity of relative entropy, and determines the first order variation of entanglement entropy under state perturbations. Its most interesting application is arguably given in Refs.~\cite{Lashkari:2013koa,Faulkner:2013ica}, where it plays a crucial role in deriving the bulk linearized Einstein's equations about a perturbed AdS background from boundary entanglement correlations of the CFT. 

Motivated by extended black hole thermodynamics \cite{Kastor:2009wy,Dolan:2010ha,Kubiznak:2016qmn}, where the cosmological constant~$\Lambda$ is interpreted as a thermodynamic pressure $p\equiv -\Lambda/8\pi G$, an extension of the first law of entanglement was proposed in Ref. \cite{Kastor:2014dra}, which includes not only variations of the state but also of the CFT itself. It can be written as
\begin{equation}\label{eq:ext}
\delta S_{EE}=\delta \langle K_B \rangle+
  \frac{S_{EE}}{a_d^\ast}
  \delta a_d^\ast\ ,
\end{equation}
where $S_{EE}$ is the vacuum entanglement entropy associated to a ball in Minkowski and $K_B$ its modular hamiltonian. The constant $a_d^\ast$ is defined for an arbitrary CFT as
\begin{equation}\label{eq:32}
a_d^\ast=
  \begin{cases} 
  \qquad \qquad \,\,\,\,\,
  A_d
  \qquad \quad \,\,\,\, 
  \ ,
  & {\rm for\,\,d\,\,even} \vspace{6pt}\\
  \,\,(-1)^{\frac{d-1}{2}}\ln[
  Z(S^d)]/2\pi\ ,
  &  {\rm for\,\,d\,\,odd}\ . \\
 \end{cases}
\end{equation}
Here $A_d$ is the coefficient in the trace anomaly proportional to Euler's density, while for odd dimensions $a_d^\ast$ is determined by the partition function of the CFT placed on a unit sphere~$S^{d}$ (see Ref. \cite{Pufu:2016zxm} for some examples in free theories). Since $a_d^\ast$ has a monotonous behavior under renormalization group flows \cite{Casini:2017vbe}, we can interpret it as counting the number of degrees of freedom in the CFT. The generalized central charge $a^{\ast}_{d}$ has appeared in a number of holographic $c$-theorems in arbitrary dimensions and higher curvature theories of gravity~\cite{Myers:2010tj}. 

The first term in (\ref{eq:ext}) is the ordinary contribution to the first law obtained by perturbing the state, while the second gives the behavior of the entanglement entropy when varying the CFT. We must emphasize that this second contribution is \textit{not} equivalent to a renormalization group flow, since the variation continuously interpolates between CFTs. It simply gives the dependence of the entanglement entropy on the CFT data.

The extended first law (\ref{eq:ext}) was initially derived in Ref. \cite{Kastor:2014dra} for a holographic CFT dual to Einstein gravity, and later generalized to specific higher curvature gravity theories in Refs. \cite{Kastor:2016bph,Caceres:2016xjz,Lan:2017xcl}. These derivations start by considering a particular Killing horizon in pure AdS and deriving an extended bulk first law which considers variations of the cosmological constant, using either Hamiltonian perturbation theory \cite{Kastor:2016bph} or the Iyer-Wald formalism \cite{Caceres:2016xjz}. The horizon entropy associated to this Killing horizon is then identified as the entanglement entropy of the boundary CFT, while the variation of the cosmological constant maps to changing the generalized central charge $a^{\ast}_{d}$. 

Given the importance and wide range of applications of the first law of entanglement, we should take any reasonable generalization seriously, as it has the potential of providing new insights into the structure of space-time and entanglement in QFTs. In this work we explore the extended first law of entanglement (\ref{eq:ext}) by generalizing previous derivations to include arbitrary theories of gravity, clarifying some of its subtle features and studying its low dimensional limit.  

%\vspace{-4mm}

The outline of this article is as follows.  We start in section \ref{sec:2} by showing that a remarkably simple argument allows us to derive the bulk analog of (\ref{eq:ext}) for perturbations of any Killing horizon in pure AdS. Contrary to previous derivations, our computation is novel in its simplicity and the fact that it holds for arbitrary bulk gravity theories and Killing horizons in pure AdS, finding no need to resort to technical calculations as in Refs. \cite{Kastor:2014dra,Kastor:2016bph,Caceres:2016xjz,Lan:2017xcl}. We discuss how each of the bulk quantities is mapped to the boundary CFT, carefully analyzing some subtleties previously overlooked. Applying our construction to certain bulk Killing horizons, we derive the extended first law (\ref{eq:ext}) for the vacuum state of a CFT reduced to the following regions: a ball and the half-space in Minkowski, a spherical cap in the Lorentzian cylinder $\mathbb{R}\times S^{d-1}$ and de Sitter, and a ball in AdS$_d$. The method used to find the appropriate bulk Killing horizons crucially relies on the freedom to choose conformal frames at the AdS boundary.

%\vspace{-4mm}

We continue in section \ref{sec:2Dlaw}, where we revisit the calculations from  section \ref{sec:2} but carefully analyzing the case in which the bulk theory is two-dimensional. While for a class of two-dimensional gravity theories we find no obstructions when deriving the extended first law for Killing horizons in pure AdS$_2$, there are certain Einstein-dilaton theories where the end result takes a different form. We illustrate this in subsection \ref{sub:JT} for Jackiw-Teitelboim gravity \cite{Teitelboim:1983ux,Jackiw:1984je}, where we show the extended first law for Killing horizons is distinct.

In section \ref{sec:3Dlaw} we show that in three dimensional gravity an extended first law can be derived for Killing horizons in space-times that are locally but not globally AdS. This allows us to obtain an extended first law for the boundary CFT$_2$ that is analogous to (\ref{eq:ext}) but involving thermal instead of entanglement entropy. From the bulk perspective we find some interesting results for  extended black hole thermodynamics, where we obtain a curious formula for the thermodynamic volume (see Eq. (\ref{eq:36})), the conjugate variable to the pressure $p$.

%\vspace{-4mm}

We conclude in section \ref{sec:4} by expanding some discussions on the calculations in the main text. We clarify some aspects regarding the structure of divergences in the extended first law of entanglement (\ref{eq:ext}) and critically analyze the extent to which it can hold for arbitrary regions and CFTs. We briefly comment on the bulk constraints implied by assuming both the RT holographic entropy formula \cite{Ryu:2006bv} and the extended first law of entanglement hold for arbitrary setups in the boundary CFT. Finally, we discuss some interesting aspects of the thermodynamic volume in three dimensional gravity and its connection to the microscopic interpretation of black hole super-entropicity \cite{Cvetic:2010jb}.

%%%%%%%%%%%%%%%%%%%%%%%%%%%%%%%%%%%%%%%%%%%%%
\section{Killing horizons in pure AdS and extended first law}
\label{sec:2}
%%%%%%%%%%%%%%%%%%%%%%%%%%%%%%%%%%%%%%%%%%

In this section we present a derivation of the extended first law of entanglement for holographic CFTs described by arbitrary covariant theories of gravity in the bulk
\begin{equation}\label{genaction}
I[\lambda_i,g_{\mu \nu}]=
  \int d^{d+1}x\,\sqrt{-g}\,
  \mathcal{L}\left(g_{\mu \nu},
  \mathcal{R}_{\mu \nu\rho \sigma},\nabla_\lambda \mathcal{R}_{\mu \nu\rho \sigma},\dots
  \right),
\end{equation}
where $\mathcal{R}_{\mu \nu \rho \sigma}$ is the Riemann tensor. Each theory is characterized by a family of coupling constants $\left\lbrace \lambda_i \right\rbrace$ that are chosen such that the action admits a pure AdS vacuum solution of radius $L$. This length scale is a non-trivial function of the coupling constants of the theory~$L=L(\lambda_i)$, and the pure AdS metric only depends on $\left\lbrace \lambda_i \right\rbrace$ through $L$. A concrete and simple example of a higher curvature theory is Einstein Gauss-Bonnet gravity. Although we could also add some matter to the action, for the most part we consider pure gravity and set matter fields to zero.

Consider a Killing vector $\xi^\mu$ of the pure AdS metric $g^{\rm AdS}_{\mu \nu}(L)$ which is time-like over some region
\begin{equation}\label{eq:15}
\xi^2\equiv g_{\mu \nu}^{\rm AdS}
  \xi^\mu \xi^\nu\le 0
  \qquad \Longleftrightarrow \qquad
  {\rm Some\,\,region\,\,of\,\,AdS}\ .
\end{equation}
The surface in which the vector vanishes defines a Killing horizon. One of the central quantities characterizing this horizon is its entropy, that for an arbitrary theory is computed from Wald's functional according to \cite{Wald:1993nt,Iyer:1994ys}
\begin{equation}\label{Waldent}
S_{\xi}
  \left[
  g_{\mu \nu}^{\rm AdS}(L),\lambda_i
  \right]=
  -2\pi \int
  dV
  \left[
  \frac{\delta \mathcal{L}}
  {\delta \mathcal{R}^{\mu \nu}_{\,\,\,\,\,\,\rho \sigma}}
  n^{\mu \nu}
  n_{\rho \sigma}
  \right],
\end{equation}
where the integral is over the bifurcation Killing surface with induced volume element $dV$. The anti-symmetric tensor $n^{\mu \nu}$ is the binormal to the horizon normalized so that $n^{\mu\nu}n_{\mu \nu}=-2$. Our aim is to study the behavior of this entropy functional under general perturbations and to determine its consequences for the boundary CFT.

Let us start by considering the behavior of the entropy under metric perturbations $g_{\mu \nu}^{\rm AdS}(L)\rightarrow  g_{\mu \nu}^{\rm AdS}(L)+\delta g_{\mu \nu}$.\footnote{The perturbation $\delta g_{\mu \nu}$ can be any metric which satisfies the equations of motion obtained from (\ref{genaction}) linearized around pure AdS.} Since we are working with a Killing horizon we can apply the same methods used to study black hole thermodynamics. As computed in Ref. \cite{Iyer:1994ys} in the context of black holes, the first order variation of~(\ref{Waldent}) is determined by the Noether charge $Q_{\xi}$. This can be related to the energy $E_\xi$ associated with the Killing vector $\xi$ measured at asymptotic spatial infinity according to \cite{Iyer:1994ys}\footnote{The Iyer and Wald formalism derives the ordinary first law from a $(d-1)$-form $\chi=\delta Q_{\xi}-\xi\cdot\Theta$ that is closed $d\chi=0$ on shell, where $\Theta$ is the symplectic potential and $\delta Q_{\xi}$ is variation of the Noether charge associated to $\xi$. The integral of $\chi$ vanishes on-shell so that one obtains a relation between the two boundary contributions, at the bifurcate Killing horizon and the asymptotic boundary. At the horizon we have $\xi=0$ so that $\chi=\delta Q_{\xi}$, and $\int_{\text{horizon}}\chi =\frac{\kappa}{2\pi}\delta S_{\xi}$, while at the asymptotic boundary $\int_{\infty}\chi=\delta E_{\xi}$ (see also Ref. \cite{Faulkner:2013ica}).}
\begin{equation}\label{Waldvar1}
\delta S_\xi=
\frac{2\pi}{\kappa}
  \delta E_\xi\ ,
  \qquad \qquad \quad
  \kappa^2=-\frac{1}{2}\left(\nabla^\mu \xi^\nu\right)\left(\nabla_\mu \xi_\nu\right)\ ,
\end{equation}
where $\kappa$ is the surface gravity.

We now consider another type of perturbation obtained by changing the gravitational theory itself, \textit{i.e.} $\mathcal{L}\rightarrow \mathcal{L}+\delta \mathcal{L}$, implemented by slightly changing the coupling constants of the theory $\lambda_i\rightarrow \lambda_i+\delta \lambda_i$. Since the pure AdS metric $g_{\mu \nu}^{\rm AdS}(L)$ is a function of $\lambda_i$ through~${L=L(\lambda_i)}$, the perturbation induces a variation of the metric. If we did not take this metric variation into account, the perturbed metric would not be a solution of the perturbed Lagrangian. Hence, the first order variation of Wald's functional is explicitly given by
\begin{equation}\label{eq:Waldvar2}
\delta S_\xi=
  S_\xi\left[g_{\mu \nu}^{\rm AdS}(\lambda_i+\delta \lambda_i),
  \lambda_i+\delta \lambda_i\right]-
  S_\xi\left[g_{\mu \nu}^{\rm AdS}(\lambda_i),
  \lambda_i\right].
\end{equation}

From the definition of Wald's entropy in (\ref{Waldent}) we can compute this in full generality, the key feature being that both terms are evaluated in the pure AdS metric of each theory. Since AdS is maximally symmetric, the integrand in (\ref{Waldent}) can be evaluated explicitly \cite{Myers:2010tj} and written as\footnote{To obtain this general expression the only thing that is required is that the metric is locally AdS, see section 5.2 of Ref. \cite{Myers:2010tj} for details. This becomes very useful in Sec. \ref{sec:3Dlaw}, where it allows us to extend some of our results beyond pure AdS in three dimensional gravity.}
\begin{equation}\label{eq:24}
\left.
\frac{\delta \mathcal{L}}
{\delta \mathcal{R}^{\mu \nu}_{\quad \rho \sigma}}
\right|_{\rm AdS}=
-\frac{L^2}{4d}
 \left(
 \delta_{\mu}^{\rho}
 \delta_{\nu}^{\sigma}-
 \delta_{\mu}^{\sigma}
 \delta_{\nu}^{\rho}
 \right)
 \left.
 \mathcal{L}
 \right|_{\rm AdS}\ ,
\end{equation}
where $\mathcal{L}\big|_{\rm AdS}$ is the Lagrangian density (\ref{genaction}) evaluated in the pure AdS solution. Using this, we can evaluate Wald's functional and write it as
\begin{equation}\label{eq:2}
S_{\xi}
  \left[
  g_{\mu \nu}^{\rm AdS}(\lambda_i),\lambda_i
  \right]=
  \frac{4\pi a_d^\ast(\lambda_i)}
  {{\rm Vol}(S^{d-1})}
  \widetilde{\mathcal{A}}_{\rm horizon}
  \ ,
\end{equation}
where $\widetilde{\mathcal{A}}_{\rm horizon}$ is the horizon area $\mathcal{A}_{\rm horizon}$ divided by the AdS radius $L^{d-1}$. We have identified~$a_d^\ast$ according to \cite{Myers:2010tj,Casini:2011kv}
\begin{equation}\label{eq:9}
a_d^\ast(\lambda_i)=-
  \frac{1}{2d}
  {\rm Vol}(S^{d-1})L^{d+1}
  \mathcal{L}
  \big|_{\rm AdS}\ ,
\end{equation}
where ${\rm Vol}(S^{d-1})=2\pi^{d/2}/\Gamma(d/2)$. The coefficient $a_d^\ast=a_d^\ast(\lambda_i)$ is in general a complicated function of the coupling constants of the theory. Using (\ref{eq:2}) we can easily evaluate the variation in (\ref{eq:Waldvar2}) and find
\begin{equation}\label{Waldvar3}
\delta_{\lambda_i} S_\xi=
  \frac{S_\xi}{a_d^\ast}\delta_{\lambda_i} a_d^\ast\ ,
  \qquad \qquad
  \delta_{\lambda_i} a_d^\ast(\lambda_i)=\sum_i
  \left(
  \frac{\partial a_d^\ast}{\partial \lambda_i}
  \right)
  \delta \lambda_i\ .
\end{equation}
This expression relies on the fact that the pure AdS metric $g_{\mu \nu}^{\rm AdS}(L)$ is only a function of the length scale ${L=L(\lambda_i)}$, which means the dimensionless horizon area ${\widetilde{\mathcal{A}}_{\rm horizon}=\mathcal{A}_{\rm horizon}/L^{d-1}}$ is independent of $\lambda_i$. In section \ref{sec:3Dlaw} we revisit this when considering more general metrics in three dimensional gravity.

Since we are considering linear perturbations we can put together the results in Eqs.~(\ref{Waldvar1}) and (\ref{Waldvar3}) and obtain the following bulk extended first law
\begin{equation}\label{bulkextlaw}
\delta S_\xi=
  \frac{2\pi}{\kappa}\delta E_\xi'+
  \frac{S_\xi}{a_d^\ast}\delta a_d^\ast\ .
\end{equation}
We can already see the similarities of this bulk relation with the extended first law of entanglement (\ref{eq:ext}). For a particular Killing vector $\xi$ in AdS, this result was first obtained in Ref. \cite{Kastor:2014dra} for Einstein gravity and later in Refs. \cite{Kastor:2016bph,Caceres:2016xjz,Lan:2017xcl} for specific higher curvature gravity theories.\footnote{In some of these papers this relation is not written in terms of the coefficient $a_d^\ast$, but in terms of the coupling constants $\left\lbrace \lambda_i \right\rbrace$ of particular theories.} Our derivation generalizes to arbitrary covariant theories of gravity as well as any Killing horizon in pure AdS. The method is quite simple and follows almost immediately upon evaluating Wald's functional in (\ref{eq:2}).\footnote{Comparing with the methods in \cite{Caceres:2016xjz,Lan:2017xcl} we find we do not have to explicitly deal with additional divergences that arise (and ultimately cancel) from evaluating the Iyer-Wald form at the asymptotic boundary when implementing the extended Iyer-Wald formalism (see also \cite{Jacobson:2018ahi}).}

Before analyzing the holographic consequences of this relation, let us comment on the prime we have added on the charge $E_\xi'$ in (\ref{bulkextlaw}). From the derivation of (\ref{Waldvar3}) it is clear that when the variation is only given by ${\lambda_i\rightarrow \lambda_i+\delta \lambda_i}$, the first term in (\ref{bulkextlaw}) vanishes, $\delta_{\lambda_i} E'_{\xi}=0$, \textit{i.e.} 
\begin{equation}\label{eq:16}
E_\xi\left[
  g_{\mu \nu}^{\rm AdS}(\lambda_i+\delta\lambda_i),\lambda_i+\delta \lambda_i
  \right]-
  E_\xi\left[
  g_{\mu \nu}^{\rm AdS}(\lambda_i),\lambda_i
  \right]=0\ .
\end{equation}
Given that there is no reason for these terms to cancel each other for arbitrary values of $\lambda_i$, both must vanish separately. This is achieved by defining the normalized quantity $E_\xi'$  as
\begin{equation}\label{eq:18}
E'_\xi\left[
  g_{\mu \nu},\lambda_i
  \right]=
  E_\xi\left[
  g_{\mu \nu},\lambda_i
  \right]-
  E_\xi\left[
  g_{\mu \nu}^{\rm AdS}(\lambda_i),
  \lambda_i
  \right]\ .
\end{equation}
While this normalization plays no role in (\ref{Waldvar1}) when considering metric perturbations, it gives the appropriate behavior under more general variations. This prescription is equivalent to subtracting the Casimir energy contribution in pure AdS, that is present for certain foliations of the space-time (see Ref. \cite{Emparan:1999pm} for some examples). The procedure is common in extended black hole thermodynamics, where the Casimir energy is not included in the first law \cite{Kastor:2009wy}.

%%%%%%%%%%%%%%%%%%%%%%%%%%%%%%%%%%%%%%%%%%
\subsection{Mapping to boundary CFT}
%%%%%%%%%%%%%%%%%%%%%%%%%%%%%%%%%%%%%%%%%%

We are mainly interested in the first law in (\ref{bulkextlaw}) from the perspective of a holographic CFT. Taking a bulk coordinate $z$ so that the AdS boundary is located at $z\rightarrow 0$, the $d$-dimensional space-time in which the CFT is defined is given by
\begin{equation}\label{eq:33}
\lim_{z\rightarrow 0}ds_{\rm bulk}^2=
  w^2(x^\mu)ds^2_{\rm CFT}+\dots\ .
\end{equation}
Applying a bulk diffeomorphism or changing the definition of $w^2(x^\mu)$ results in a different boundary space-time. A particular way of taking this limit corresponds to choosing a conformal frame. We will shortly take advantage of this freedom, which from the CFT perspective is equivalent to a conformal transformation.

What about the quantum state of the boundary CFT? Although the bulk space-time is pure AdS, the CFT is technically not in the vacuum state since there is a horizon and therefore an associated temperature, given by the surface gravity in (\ref{Waldvar1}) according to $\beta=2\pi/\kappa$. This means the boundary state is thermal with respect to the Killing flow evaluated at the boundary, \textit{i.e.}
\begin{equation}\label{cftstate}
\rho=\frac{1}{Z}\exp\left(-\beta K_{\xi}\right)\ ,
\end{equation}
where the operator $K_{ \xi}$ generates the flow of $\xi^\mu$ as we approach the boundary. It can be written explicitly in terms of the boundary coordinates $x^a$ and the pullback of the Killing vector $\xi^a$ as
\begin{equation}\label{modHam}
K_{\xi}=\int_{\Sigma_\xi} 
   \xi^a T_{ab}dS^b\ ,
\end{equation}
where $T_{ab}$ is the stress tensor of the CFT and the integral is over a boundary codimension one space-like surface $\Sigma_\xi$ where the vector $\xi^a$ is time-like. The directed surface element~$dS^a$ is given by $dS^a=dSn^a$, with $n^a$ a unit vector normal to $\Sigma_\xi$. 

The variation of the conserved quantity $E_\xi'$ appearing in the gravitational first law (\ref{bulkextlaw}) is given by the variation of the expectation value of $K_{ \xi}$ in the state (\ref{cftstate}). The normalization condition for $E_\xi'$ in (\ref{eq:18}) translates into the following normalization of the stress tensor $T_{ab}$
\begin{equation}\label{eq:20}
T_{ab}\quad \rightarrow \quad T_{ab}'=
  T_{ab}-\langle T_{ab} \rangle_{\rho}\ ,
\end{equation}
with $\rho$ in (\ref{cftstate}). Since a bulk Killing vector gives a \textit{conformal} Killing vector at the boundary, the operator $K_{\xi}$ does not correspond to the Hamiltonian in general. We shall shortly consider some examples which illustrate this.

Putting everything together, the gravitational first law (\ref{bulkextlaw}) maps to the boundary CFT according to
\begin{equation}\label{extlawent}
\delta S=
  \beta\,\delta \langle K_{\xi} \rangle_{\rho}+
  \frac{S}{a_d^\ast}\delta a_d^\ast\ ,
\end{equation}
where we identified the horizon entropy $S_\xi$ with the Von Neumann entropy ${S(\rho)=-{\rm Tr}(\rho\ln(\rho))}$ of $\rho$ in (\ref{cftstate}). From the field theory perspective it might not be entirely clear what each of these terms corresponds to, so let us write them more explicitly.

For perturbations in which we keep the CFT fixed it is clear that $\delta a_d^\ast=0$ while the state is deformed according to $\rho+\delta \rho$. In this case, the relation (\ref{extlawent}) is similar to the first law of thermodynamics. When ${\delta a_d^\ast\neq 0}$ we must be more careful since in this case the CFT is changing, which in particular implies that the Hilbert space shifts $\mathcal{H}\rightarrow \bar{\mathcal{H}}$. The state $\rho$ cannot remain fixed, meaning that $\delta a_d^\ast \neq 0$ induces a variation of $\rho$ given by 
\begin{equation}
\rho \qquad \longrightarrow \qquad
  \bar{\rho}=\frac{1}{Z}\exp\left(
  -\beta \bar{K}_{\xi}
  \right)\ ,
\end{equation}
where $\bar{\rho}$ and $\bar{K}_{\xi}$ are the same operators but acting on the Hilbert space $\bar{\mathcal{H}}$ instead. In this case the extended first law (\ref{extlawent}) can be written explicitly as
\begin{equation}\label{eq:21}
S(\bar{\rho})-S(\rho)=
  \beta\left[
  \langle \bar{K}_{ \xi}
  \rangle_{\bar{ \rho}}-
  \langle 
  K_{ \xi}
  \rangle_\rho
  \right]+
  \frac{S(\rho)}{a_d^\ast}
  \delta a_d^\ast\ .
\end{equation}
Notice that the first terms on the right-hand side involve operators on different Hilbert spaces. Moreover, the normalization of $K_{\xi}$ given in (\ref{eq:20}) (and an analogous expression for $\bar{K}_{\xi}$) implies that both terms between square brackets vanish independently. This is equivalent to the gravitational case, where we obtained (\ref{Waldvar3}).

Putting everything together, the most general perturbation of the Von Neumann entropy of $\rho$ is given by
\begin{equation}\label{eq:12}
S(\bar{\rho}+\delta \bar{\rho})-
  S(\rho)=
  \beta\,
  {\rm Tr}\left(
  \bar{K}_\xi \,\delta \bar{\rho}
  \right)
  +
  \frac{S(\rho)}{a_d^\ast}
  \delta a_d^\ast\ ,
\end{equation}
where we have used $\langle K_{ \xi}\rangle_\rho=\langle \bar{K}_{ \xi}
  \rangle_{\bar{\rho}}=0$. This expression considers the simultaneous variations~$a_d^\ast\rightarrow a_d^\ast+\delta a_d^\ast$ and $\rho \rightarrow\bar{\rho}+\delta \bar{\rho}$, and clarifies the precise meaning of (\ref{extlawent}), which without any explanation is rather obscure.

%%%%%%%%%%%%%%%%%%%%%%%%%%%%%%%%%%%%%%
\subsection{Extended first law of entanglement}
\label{sec:3}
%%%%%%%%%%%%%%%%%%%%%%%%%%%%%%%%%%%%%%

So far we have shown that (\ref{extlawent}) follows from AdS/CFT when studying Killing horizons in pure AdS. We now consider particular horizons that will allow us to identify this relation as the extended first law of entanglement. Let us start with the simplest example of a Killing horizon in AdS, obtained by writing pure AdS in a hyperbolic slicing
\begin{equation}\label{adshyp}
ds^2=-\left(\frac{\rho^2-L^2}{R^2}\right)d\tau^2+
  \left(\frac{L^2}{\rho^2-L^2}\right)d\rho^2+
  \rho^2dH^2_{d-1}\ ,
\end{equation}
where $R$ is an arbitrary positive constant and $dH_{d-1}$ is the line element of a unit hyperbolic plane. This space-time is often referred as Rindler-AdS since it describes a section of anti-de Sitter. The vector $\xi=\partial_\tau$ trivially satisfies Killing's equation and is time-like over the whole patch $\rho\ge L$, generating a horizon at $\rho=L$. It therefore satisfies all the conditions leading to the first law in (\ref{bulkextlaw}) and (\ref{eq:12}). 

A simple computation shows that the surface gravity is $\kappa=1/R$, while the boundary metric is given by ${ds^2_{\rm CFT}=-d\tau^2+R^2dH^2_{d-1}\equiv \mathbb{R}\times \mathbb{H}^{d-1}}$. From this we see that $\xi=\partial_\tau$ is also a Killing vector of $ds^2_{\rm CFT}$, so that $K_\xi$ in (\ref{modHam}) is equal to the Hamiltonian and can be written as
\beq K_\xi=
  \int_{\tau=0}
  T_{\tau \tau}'dS^\tau \equiv
  H_\tau\ .\eeq
This means the boundary state is an ordinary thermal state ${\rho_\beta\propto \exp(-\beta H_\tau)}$, where the inverse temperature is fixed by the surface gravity to ${\beta=2\pi R}$. The extended first law (\ref{extlawent}) then becomes
\begin{equation}\label{eq:55}
\delta S(\rho_\beta)=\beta \,
  \delta \langle 
  H_\tau
  \rangle+\frac{S(\rho_\beta)}{a_d^\ast}\delta a_d^\ast\ .
\end{equation}
While the first term is nothing more than the first law of thermodynamics, the second contribution is unique to the case of inverse temperature $\beta=2\pi R$. This is clear from the holographic perspective, since moving away from this temperature is equivalent to leaving pure AdS, where the analysis of the previous section is no longer valid. In section \ref{sec:3Dlaw} we show that for $d=2$ this expression remains valid for arbitrary values of $\beta$. Although (\ref{eq:55}) is not the extended first law of entanglement (since it involves a thermal state in $\mathbb{R}\times \mathbb{H}^{d-1}$), this simple example will be very useful in what follows.

%%%%%%%%%%%%%%%%%%%%%%%%%%%%%%%%%%%%%%%%%%%%%
\subsubsection{Shifting conformal frames}
%%%%%%%%%%%%%%%%%%%%%%%%%%%%%%%%%%%%%%%%%%

Building on the canonical example we just described, we can obtain the more complicated setups we are actually interested in. To obtain the extended first law of entanglement we take advantage of the freedom present when taking the boundary limit in (\ref{eq:33}). Different ways of taking this limit correspond to distinct conformal frames and result in different setups for the boundary CFT. We still consider the bulk Killing vector $\xi=\partial_\tau$, but written in a different set of coordinates corresponding to distinct conformal frames.

%%%%%%%%%%%%%%%%%%%%%%%%%%%%%55
\subsubsection*{Ball in Minkowski}
%%%%%%%%%%%%%%%%%%%%%%%%%%%%%%%

Let us first show how we can recover the extended first law of entanglement for the Minkowski vacuum reduced to a ball. We first apply a change of coordinates on the Rindler-AdS metric~(\ref{adshyp}), which is given in Eq. (4.7) of Ref.~\cite{Rosso:2019lsm}, so that the metric becomes
\begin{equation}\label{eq:26}
ds^2=
  \left(\frac{L}{\hat{r}\sin(\psi)}\right)^2
  \left(
  -dt^2+d\hat{r}^2+
  \hat{r}^2\left(d\psi^2+
  \cos^2(\psi)d\Omega^2_{d-2}\right)
  \right),
\end{equation}
where $\hat{r}\ge 0$, $\psi\in [0,\pi/2]$ and $d\Omega_{d-2}$ is the line element of a unit sphere $S^{d-2}$. This is nothing more than the AdS Poincar\'e patch, as can be seen by defining the usual coordinates~${(z,r)=\hat{r}\,(\sin(\psi),\cos(\psi))}$. At the boundary $\psi \rightarrow 0$ we recover $d$-dimensional Minkowski space-time with $\hat{r}=r$ the spatial radial coordinate. We use the convention in which the boundary coordinate $r$ refers to the bulk coordinate $\hat{r}$ when $\psi\rightarrow 0$. This same notation is used in the following examples.

It is straightforward to write the Killing vector $\xi=\partial_\tau$ in these new coordinates and find
\begin{equation}\label{eq:23}
\xi=\left(\frac{R^2-\hat{r}_+^2}{2R^2}\right)
  \partial_{\hat{r}_+}-
  \left(\frac{R^2-\hat{r}_-^2}{2R^2}\right)
  \partial_{\hat{r}_-}\ ,
\end{equation}
where $\hat{r}_\pm=\hat{r}\pm t$. The important difference with respect to the hyperbolic example is that this Killing vector is time-like only in a section of the metric (\ref{eq:26}), given by $|\hat{r}_\pm|\le R$. For the Minkowski boundary this corresponds to the causal domain of a ball of radius $R$. The operator generating the flow of $\xi$ inside the ball can be written from (\ref{modHam}) as
\begin{equation}\label{eq:22}
K_{\xi}=
  \int_{r\le R}
  \left(
  \frac{R^2-r^2}{2R^2}
  \right)T'_{tt}\,
  dS^t\ .
\end{equation}

While this is clearly not the Hamiltonian generating $t$ translations in Minkowski, it is proportional to the modular hamiltonian characterizing the Minkowski vacuum reduced to the ball \cite{Casini:2011kv}. The proportionality constant missing to make the identification is given by~${K_{\rm Ball}=2\pi R K_\xi}$, that is precisely the inverse temperature $\beta=2\pi R$ obtained from the surface gravity of the bulk Killing vector (\ref{eq:23}). Altogether, the quantum state $\rho$ in (\ref{cftstate}) is exactly given by the Minkowski vacuum reduced to the ball. The Von Neumann entropy is equivalent to the entanglement entropy, so that (\ref{extlawent}) becomes the extended first law of entanglement (\ref{eq:ext}).

%%%%%%%%%%%%%%%%%%%%%%%%%%%%%%%%%%%%%%
\subsubsection*{Half-space in Minkowski}
%%%%%%%%%%%%%%%%%%%%%%%%%%%%%%%%%%%%%

Another interesting case is obtained by applying the change of coordinates given in Eq. (4.4) of Ref. \cite{Rosso:2019lsm} (see also Ref. \cite{Emparan:1999gf}) to the Rindler-AdS space-time, so that the bulk metric (\ref{adshyp}) becomes
\begin{equation}
ds^2=\left(L/z\right)^2
  \left(
  dz^2-dt^2+dx^2+d\vec{y}.d\vec{y}\,
  \right)\ ,
\end{equation}
where $(x,\vec{y}\,)\in \mathbb{R}\times \mathbb{R}^{d-2}$. Once again we recognize the Poincar\'e patch of AdS, so that we recover a $d$-dimensional Minkowski boundary when $z\rightarrow 0$. The Killing vector $\xi=\partial_\tau$ in these coordinates is given by
\begin{equation}\label{eq:25}
\xi=
  (x_+/R)\partial_{x_+}-
  (x_-/R)\partial_{x_-}\ ,
\end{equation}
where $x_\pm=x\pm t$. This vector is time-like when $x_\pm \ge 0$, which from the boundary perspective corresponds to the Rindler region, \textit{i.e.} the causal domain of the half space $x\ge 0$. Using (\ref{modHam}) to compute the operator generating the Killing flow at the boundary we find
\beq K_\xi=
  \int_{x>0}(x/R)T_{tt}'\,dS^t\ .\eeq
Since the surface gravity of (\ref{eq:25}) is still given by $\kappa=1/R$, the inverse temperature is~${\beta=2\pi R}$ and we recognize ${\rho\propto \exp(-\beta K_\xi)}$ as the Minkowski vacuum reduced to Rindler~\cite{Bisognano:1976za,Unruh:1976db}. Similarly to the previous case, (\ref{extlawent}) becomes the extended first law of entanglement (\ref{eq:ext}) but in this case, for the Minkowski vacuum reduced to the half-space.

%%%%%%%%%%%%%%%%%%%%%%%%%%%%%%%%%%%%
\subsubsection*{Spherical cap in Lorentzian cylinder}
%%%%%%%%%%%%%%%%%%%%%%%%%%%%%%%%%%%%%

Let us now show how we can obtain the extended first law of entanglement for holographic CFTs defined on curved backgrounds. Consider the following change of coordinates on the AdS metric (\ref{eq:26})
\begin{equation}\label{eq:27}
\hat{r}_\pm(\hat{\theta}_\pm)=R\frac{\tan(\hat{\theta}_\pm/2)}{\tan(\theta_0/2)}\ ,
\end{equation}
where $\hat{\theta}_\pm=\hat{\theta}\pm \sigma/R$ and $\theta_0\in[0,\pi]$ is a fixed parameter. The metric (\ref{eq:26}) becomes
\begin{equation}\label{eq:28}
ds^2=\left[
  \frac{L/R}{\sin(\psi)\sin(\hat{\theta})}
  \right]^2 
  \left(
  -d\sigma^2
  +R^2d\hat{\theta}^2
  \right.  \left.
  + R^2\sin^2(\hat{\theta})
  \left(
  d\psi^2
  +\cos^2(\psi)d\Omega^2_{d-2}
  \right)
  \right)\ ,
\end{equation}
where $\sigma\in \mathbb{R}$ is the time coordinate and $\hat{\theta}$ is restricted to $\hat{\theta}\in[0,\pi]$. As we take the boundary limit $\psi\rightarrow 0$ and remove the conformal factor between square brackets we find that the CFT is defined in the Lorentzian cylinder $\mathbb{R}\times S^{d-1}$ with metric~${ds^2_{\rm CFT}=-d\sigma^2+R^2d\Omega^2_{d-1}}$. The bulk coordinate $\hat{\theta}$ becomes the polar angle $\hat{\theta}=\theta$ on the spatial sphere~$S^{d-1}$, with~$\theta=0,\pi$ corresponding to the North and South poles respectively.

\begin{figure}
\centering
\includegraphics[scale=0.80]{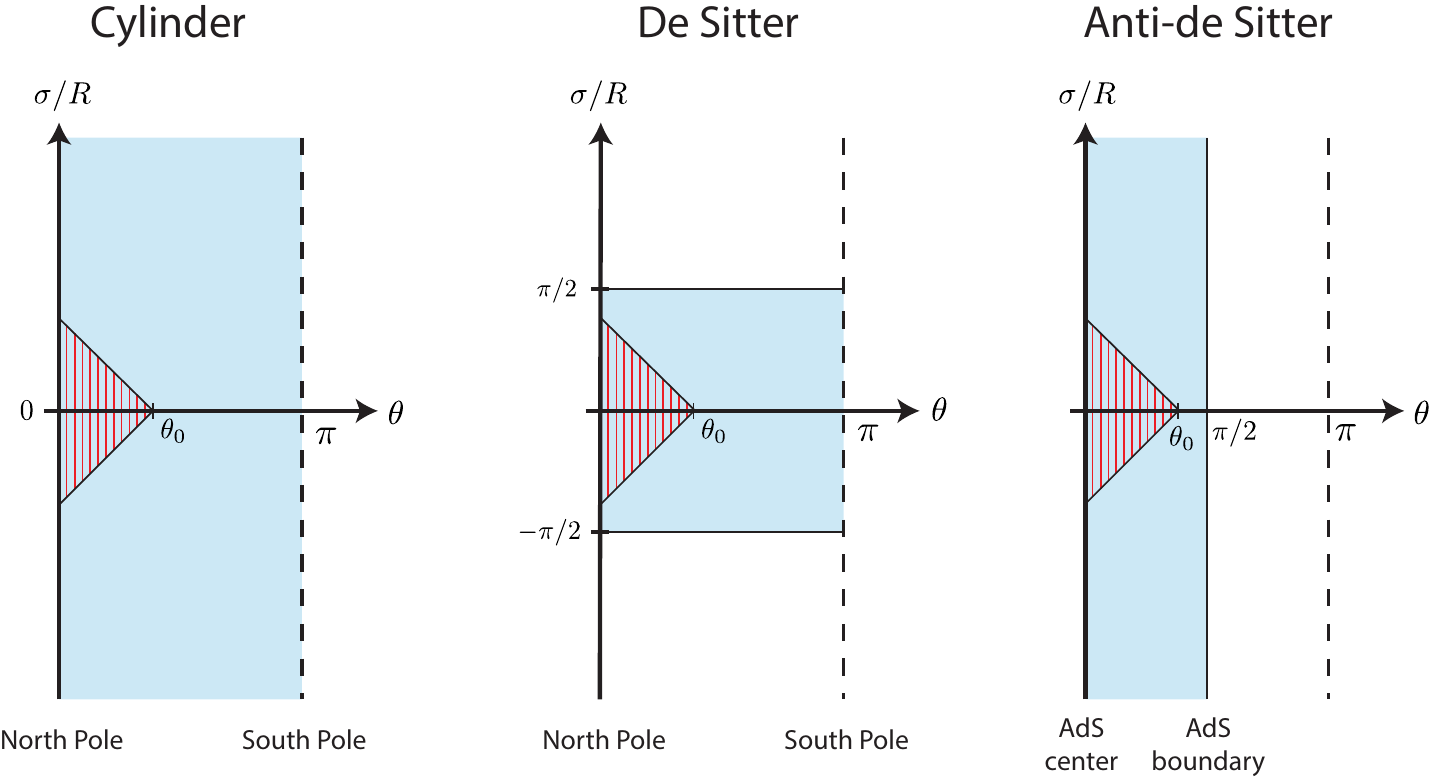}
\caption{Boundary space-times represented in the $(\sigma/R,\theta)$ plane. The blue region corresponds to the section of the $(\sigma/R,\theta)$ plane covered by the boundary metrics (\ref{eq:28}) (in the limit $\psi\rightarrow 0$ and without the conformal factor), (\ref{eq:11}) and (\ref{eq:51}). In red we see the region in which the boundary vector $\xi^a$ is time-like and therefore the extended first law of entanglement applies.}\label{fig:1}
\end{figure}

The Killing vector $\xi$ in (\ref{eq:23}) can be written in these coordinates as
\begin{equation}\label{eq:19}
\xi=
  \bigg(
  \frac{\cos(\hat{\theta}_+)-\cos(\theta_0)}
  {R\sin(\theta_0)}
  \bigg)\partial_{\hat{\theta}_+}-
  \bigg(
  \frac{\cos(\hat{\theta}_-)-\cos(\theta_0)}
  {R\sin(\theta_0)}
  \bigg)
  \partial_{\hat{\theta}_-}.
\end{equation}
Computing its magnitude we see that the bulk region in which this vector is time-like is given by $|\hat{\theta}_\pm|<\theta_0$. For the boundary CFT in the Lorentzian cylinder, this corresponds to the causal domain of a spherical cap on the spatial $S^{d-1}$ given by $\theta\in[0,\theta_0]$ at $\sigma=0$. Plotting this region in the $(\sigma/R,\theta)$ plane we obtain the left diagram in Fig. \ref{fig:1}. The whole infinite strip in blue corresponds to the Lorentzian cylinder $\mathbb{
R}\times S^{d-1}$, with the North and South pole located at $\theta=0,\pi$. 

The operator generating the Killing flow at the boundary is computed from (\ref{modHam}) as 
\begin{equation}\label{eq:29}
K_\xi=\int_{\theta\le \theta_0}
  \bigg(
  \frac{\cos(\theta)-\cos(\theta_0)}
  {R\sin(\theta_0)}
  \bigg)
  T'_{\sigma \sigma}\, dS^\sigma\ .
\end{equation}
In a similar way to the previous case, we recognize the state ${\rho \propto \exp\left(-\beta K_\xi\right)}$ with $\beta=2\pi R$ as the vacuum state of the cylinder reduced to the spherical cap \cite{Casini:2011kv}. This gives the extended first law of entanglement for a CFT in the Lorentzian cylinder (\ref{eq:ext}).

%%%%%%%%%%%%%%%%%%%%%%%%%%%%%%%%%%%%%%%%
\subsubsection*{Spherical cap in de Sitter}
%%%%%%%%%%%%%%%%%%%%%%%%%%%%%%%%%%%%%%%%

Using the same coordinates as in (\ref{eq:28}) we can obtain a CFT defined on a de Sitter background by taking the limit $\psi \rightarrow 0$ and choosing the conformal factor so that the boundary metric is given by
\begin{equation}\label{eq:11}
ds^2_{\rm CFT}=
  \frac{-d\sigma^2+R^2d\Omega^2_{d-1}}
  {\cos^2(\sigma/R)}\ .
\end{equation}
This is $d$-dimensional global de Sitter space-time, as can be seen by changing the time coordinate to $\cosh(t_s/R)=1/\cos(\sigma/R)$, so that we get
\begin{equation}\label{eq:40}
ds^2_{\rm CFT}=
  -dt_s^2+R^2\cosh^2(t_s/R)d\Omega^2_{d-1}
  \ .
\end{equation}

It is convenient to work in the time coordinate $\sigma$, since the Killing vector $\xi$ has the simple form given in (\ref{eq:19}) and is time-like when $|\theta_\pm|\le \theta_0$. Plotting this region in the~${(\sigma/R,\theta)}$ plane for the boundary metric (\ref{eq:11}), we obtain the center diagram in Fig.~\ref{fig:1}. The main difference with respect to the case of the Lorentzian cylinder is that the full de Sitter space-time (blue region) is constrained to $|\sigma/R|\le \pi/2$ due to the denominator in~(\ref{eq:11}). Since the topology of dS is the same as the cylinder $\mathbb{R}\times S^{d-1}$, the region in which $\xi^a$ is time-like also corresponds to the causal domain of a spherical cap ${\theta\in[0,\theta_0]}$, but with $\theta_0$ restricted to $\theta_0\le \pi/2$.

The operator generating the flow of the Killing vector at the boundary is still given by~(\ref{eq:29}),\footnote{The only difference with respect to the case of the cylinder is given by the induced surface element $dS^\sigma$, which is now computed from (\ref{eq:11}).} which is equivalent to the modular hamiltonian of the dS vacuum after multiplying by $\beta=2\pi R$. Altogether, this results in the extended first law of entanglement (\ref{eq:ext}) for the de Sitter vacuum reduced to a spherical cap.

%%%%%%%%%%%%%%%%%%%%%%%%%%%%%%%%%%%%%%
\subsubsection*{Ball in anti-de Sitter}
%%%%%%%%%%%%%%%%%%%%%%%%%%%%%%%%%%%%%%

Finally, we can obtain a CFT defined in an AdS$_d$ space-time by taking the limit $\psi\rightarrow 0$ in~(\ref{eq:28}) and choosing the conformal factor so that we get
\begin{equation}\label{eq:51}
ds^2_{\rm CFT}=
  \frac{-d\sigma^2+
  R^2(d\theta^2+
  \sin^2(\theta)
  d\Omega^2_{d-2})}
  {\cos^2(\theta)}
  \ .
\end{equation}
Changing coordinates to $\varrho=R\tan(\theta)\ge 0$ we recognize global AdS$_{d}$, with $\varrho$ the usual radial coordinate. Similar to the dS case, it is convenient to describe the AdS$_d$ boundary in terms of the $(\sigma,\theta)$ coordinates, where the Killing vector $\xi$ and operator $K_\xi$ are still given by (\ref{eq:19}) and (\ref{eq:29}). The main difference is that the region in which $\xi$ is time-like $|\theta_\pm|\le \theta_0$, now corresponds to the causal domain of a ball in AdS$_d$ of radius $\varrho_{\rm max}=R\tan(\theta_0)$. We plot this in the right diagram of Fig. \ref{fig:1}, where $\theta=0,\pi/2$ in (\ref{eq:51}) now correspond to the AdS center and boundary. The entanglement entropy associated to the vacuum state reduced on this ball satisfies the extended first law of entanglement in (\ref{eq:ext}).

%%%%%%%%%%%%%%%%%%%%%
\section{Killing horizons in pure AdS$_2$}
\label{sec:2Dlaw}
%%%%%%%%%%%%%%%%%%%%%%

Our calculations so far have been in the context of the AdS$_{d+1}$/CFT$_d$ correspondence for~${d\ge 2}$, where the duality is well understood. In this section we revisit the construction for the case in which $d=1$, where the gravity theory is highly constrained and there is no clear holographic picture. 

Let us start by briefly reviewing some basic notions of two dimensional gravity (see Ref.~\cite{Strobl:1999wv} for a comprehensive review). In two space-time dimensions the most general scalar curvature invariant is built from the Ricci scalar $\mathcal{R}$ and contractions of its covariant derivatives, \textit{e.g.} $(\nabla \mathcal{R})^2=(\nabla_\mu \mathcal{R})(\nabla^\mu \mathcal{R})$. Both the Riemann and Ricci tensor are fixed by $\mathcal{R}$ and~$g_{\mu \nu}$ according to
\begin{equation}\label{eq:14}
\mathcal{R}_{\mu \nu \rho \sigma}=
  \frac{\mathcal{R}}{2}
  \left(
  g_{\mu \rho}g_{\nu \sigma}-
  g_{\mu \sigma}g_{\nu \rho}
  \right)
  \ ,
  \qquad \qquad
  \mathcal{R}_{\mu \nu}=\frac{\mathcal{R}}{2}g_{\mu \nu}\ .
\end{equation}
This means there is a single gravitational degree of freedom, determined by $\mathcal{R}$. Similarly to the general $d$ case in (\ref{genaction}), the most general two dimensional gravity theory is given by
\begin{equation}\label{eq:7}
I[g_{\mu \nu},\lambda_i]=
  \int d^2x\sqrt{-g}\,
  \mathcal{L}(\mathcal{R},\nabla_\mu \mathcal{R},\dots)\ ,
\end{equation}
where the coefficients $\lambda_i$ are the coupling constants of the theory. The only constraint we impose is that there is a pure AdS solution with some radius $L=L(\lambda_i)$. Notice that the relations in (\ref{eq:14}) imply the Einstein tensor $G_{\mu \nu}=\mathcal{R}_{\mu \nu}-g_{\mu\nu}\mathcal{R}/2$ vanishes for every two dimensional metric, so that $\mathcal{L}=\mathcal{R}$ gives a trivial theory.

Just as in the higher dimensional case, let us consider a Killing vector $\xi^\mu$ of pure AdS$_2$ which is time-like over some region and generates a horizon (\ref{eq:15}). The associated entropy is computed from Wald's functional (\ref{Waldent}), that in the two dimensional case is given by
\begin{equation}\label{eq:45}
S_\xi[g_{\mu \nu}(L),\lambda_i]=
  -2\pi\left[
  \frac{\delta \mathcal{L}}
  {\delta R^{\mu \nu}_{\,\,\,\,\,\rho \sigma}}
  n^{\mu \nu}n_{\rho \sigma}
  \right]_{\rm Horizon}\ ,
\end{equation}
where there is no integral since the bifurcate horizon is a single point. Evaluating in pure AdS we can use (\ref{eq:24}) to write this as
\begin{equation}\label{eq:17}
S_\xi[g_{\mu \nu}^{\rm AdS}(L),\lambda_i]=
  2\pi a_1^*(\lambda_i)\ ,
  \qquad {\rm where} \qquad
  a_1^*(\lambda_i)=
  -L^2\mathcal{L}\big|_{\rm AdS}\ .
\end{equation}
An important difference with respect to the higher dimensional case, is that in two dimensions this expression is always finite and only depends on the global features of the theory, \textit{i.e.}, it is insensitive to the details of the Killing vector $\xi^\mu$. The entropy in (\ref{eq:17}) only depends on the pure AdS$_2$ radius and the Lagrangian density evaluated on AdS$_2$. Altogether, there is no obstruction in applying the same reasoning as in higher dimensions and write the extended first law for Killing horizons in pure AdS exactly as in (\ref{bulkextlaw})
\begin{equation}\label{eq:49}
\delta S_\xi=
  \frac{2\pi}{\kappa}
  \delta E'_\xi+
  \frac{S_\xi}{a_1^*}\delta a_1^*\ .
\end{equation}

\begin{figure}
\centering
\includegraphics[scale=0.8]{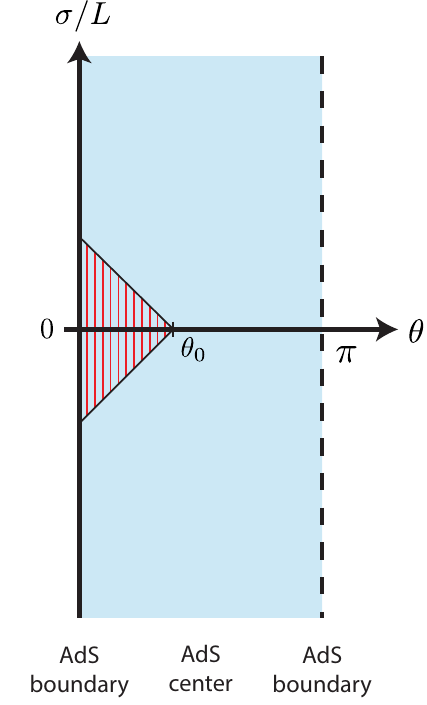}
\caption{The blue region corresponds to AdS$_2$ space-time represented in the $(\sigma/L,\theta)$ plane, with the two boundaries at $\theta=0,\pi$. In red we see the region in which the bulk Killing vector~$\xi^\mu$ (\ref{eq:52}) is time-like and therefore the extended first law in (\ref{eq:49}) applies.}\label{fig:2}
\end{figure}

Let us construct a concrete example by first writing pure AdS$_2$ in global coordinates
\begin{equation}\label{eq:13}
ds^2=
  \frac{-d\sigma^2+L^2d\theta^2}
  {\sin^2(\theta)}\ ,
\end{equation}
where $\sigma\in \mathbb{R}$ and $\theta\in(0,\pi)$. Notice this notation is different from the previous section, since $\theta$ is now a bulk coordinate and the boundary is just described by $\sigma$. Two-dimensional AdS is distinct from higher dimensions, since there are two disjoint boundaries at $\theta=0,\pi$. A sketch of its Penrose diagram is given in Fig. \ref{fig:2}.

We can easily check that the following is a Killing vector
\begin{equation}\label{eq:52}
\xi=\left(
  \frac{\cos(\theta_+)-\cos(\theta_0)}
  {L\sin(\theta_0)}
  \right)\partial_{\theta_+}-
  \left(
  \frac{\cos(\theta_-)-\cos(\theta_0)}
  {L\sin(\theta_0)}
  \right)\partial_{\theta_-}\ ,
\end{equation}
with surface gravity $\kappa=1/L$. From its norm we see that it is time-like in the domain of dependence of the bulk surface $(\sigma=0,\theta)$ with $\theta\in(0,\theta_0)$, meaning the boundary time coordinate is restricted to $|\sigma/L|\le \theta_0$. This corresponds to the red region in Fig. \ref{fig:2}.

As an example, let us compute the horizon entropy explicitly for a particular gravity theory, that we take as
\begin{equation}\label{eq:31}
\mathcal{L}=f(\mathcal{R})=
  \lambda_0+\lambda_2\mathcal{R}^2\ .
\end{equation}
The AdS radius $L$ is determined by solving the equations of motion evaluated at $\mathcal{R}=-2/L^2$, which can be written as
\begin{equation}\label{eq:30}
\nabla_\mu \nabla_\nu 
  f'(\mathcal{R})+
  \frac{1}{2}g_{\mu\nu}
  \left(
  \mathcal{R}f'(\mathcal{R})-f(\mathcal{R})
  \right)=0
  \qquad \Longrightarrow \qquad
  L^4=
  \frac{4\lambda_2}{\lambda_0}\ .
\end{equation}
Using this we can evaluate Wald's entropy in (\ref{eq:17}) as
\beq S_\xi\big[
  g_{\mu \nu}^{\rm AdS}(L),\lambda_i
  \big]=
  2\pi\left(
  -8\lambda_2/L^2
  \right)\ ,\eeq
where between parenthesis we identify the factor $a_1^*$, which is positive if and only if $\lambda_2<0$.

This raises the question regarding the holographic interpretation of the extended first law as written in (\ref{eq:49}), since $a_1^*$ is supposed to capture the number of degrees of freedom of the boundary theory. The usual AdS/CFT correspondence for a two dimensional bulk does not yield a clear picture as in the higher dimensional case. Although there has been very interesting work on the subject (see Refs. \cite{Strominger:1998yg,Cadoni:1999ja,Hartman:2008dq,Castro:2008ms,Alishahiha:2008tv,Cvetic:2016eiv}), there continues to be debate about what is meant by the dual ``$\text{CFT}_{1}$", whether it is conformal quantum mechanics or the chiral sector of a two-dimensional CFT. Moreover in the context of Jackiw-Teitelboim (JT) gravity \cite{Teitelboim:1983ux,Jackiw:1984je} it is understood that the boundary is not a single theory but an ensemble average \cite{Saad:2019lba}. For these reasons, we refrain from giving a boundary interpretation of the extended first law and leave this aspect to future investigations.

%%%%%%%%%%%%%%%%%%%%%%%%%%%%%%%%%%%%%%%
\subsection{Einstein-dilaton theories}
\label{D2GRlimit}
%%%%%%%%%%%%%%%%%%%%%%%%%%%%%%%%%%%%%%%

So far we have considered two dimensional theories of gravity in which the only field is given by the metric $g_{\mu \nu}$. We now discuss the extended first law for Einstein-dilaton theories, which are widely studied in the context of two dimensional gravity.

One disadvantage of the pure gravity action considered in (\ref{eq:7}) is that since non-trivial theories must have $\mathcal{L}\sim \mathcal{O}(\mathcal{R}^2)$, the equations of motion for the metric are at least fourth order differential equations. This issue can be avoided by the introduction of an auxiliary dilaton field $\phi(x^\mu)$ coupled to ordinary Einstein gravity
\begin{equation}\label{eq:8}
I_\phi[g_{\mu \nu},\lambda_i]=
  \int d^2x\sqrt{-g}\left[
  \phi\mathcal{R}-
  V(\phi)
  \right]\ .
\end{equation}
The equations of motion obtained from this action are second order. In particular, varying with respect to the dilaton field we get the algebraic constraint $\mathcal{R}=V'(\phi)$. If the potential has non-vanishing second derivative, one can invert this relation and substitute back into the action (\ref{eq:8}) to obtain a purely gravitational theory of the type $\mathcal{L}=f(\mathcal{R})$. As an example, if we take $V(\phi)=\phi^2/4\lambda_2-\lambda_0$, the equation of motion for $\phi$ sets $\phi_0=2\lambda_2\mathcal{R}$ and we get
\begin{equation}\label{eq:46}
I_{\phi=\phi_0}[g_{\mu\nu},\lambda_i]=
  \int d^2x\sqrt{-g}
  \left[
  \lambda_0+\lambda_2\mathcal{R}^2
  \right]\ ,
\end{equation}
that is the gravity theory previously considered in (\ref{eq:31}). This allows us to study two dimensional gravity from the simpler action (\ref{eq:8}). We should interpret the dilaton field as a gravitational degree of freedom, which gets non-trivial dynamics from varying (\ref{eq:8}) with respect to the metric
\begin{equation}\label{eq:48}
\nabla_\mu \nabla_\nu 
  \phi
  =\frac{1}{2}
  g_{\mu \nu}
  V(\phi)\ .
\end{equation} 
Since the Einstein-dilaton theories in (\ref{eq:8}) (with $V''(\phi)\neq 0$) are equivalent to the purely gravitational action previously considered in (\ref{eq:7}), the results obtained for the extended first law also hold in this setup. We should mention that while JT gravity is given by (\ref{eq:8}) with~$V(\phi)\propto \phi$, it cannot be written as a purely gravitational theory since $V''(\phi)=0$ and the dilaton equation simply fixes the curvature to a constant $\mathcal{R}={\rm const}$. We analyze the case of JT gravity separately in the next subsection.

There are more general Einstein-dilaton actions than (\ref{eq:8}) that yield interesting two dimensional theories. For instance, there is a particular way of taking the two-dimensional limit of higher dimensional Einstein gravity which results in the following action \cite{Mann:1992ar}
\begin{equation}\label{eq:47}
I_\phi\left[g_{\mu \nu},\Lambda_2\right]=
\int d^{2}x\sqrt{-g}\left[\phi\mathcal{R}+\frac{1}{2}(\nabla\phi)^{2}-2\Lambda_{2}\right]\;,
\end{equation} 
where $\Lambda_2$ is a coupling constant. This theory was studied in Ref. \cite{Frassino:2015oca} from the perspective of extended black hole thermodynamics. Although this action is clearly different from (\ref{eq:8}), if we redefine the metric according to $\tilde{g}_{\mu \nu}=e^{\phi/2}g_{\mu \nu}$ it can be written as
\begin{equation}
I_\phi\left[\tilde{g}_{\mu \nu},\Lambda_2\right]=
\int d^{2}x\sqrt{-\tilde{g}}
  \big[
  \phi\tilde{\mathcal{R}}
  -V(\phi)\big]\ ,
  \qquad {\rm where} \qquad
  V(\phi)=
  2\Lambda_{2}e^{-\phi/2}\ .
\end{equation}
Once we have the action in this form, we can solve the dilaton field equation and substitute back into the action to get a purely gravitational theory for the metric $\tilde{g}_{\mu \nu}$
\begin{equation}\label{eq:50}
I_{\phi=\phi_0}
\left[\tilde{g}_{\mu \nu},\Lambda_2\right]=
\int d^{2}x\sqrt{-\tilde{g}}
  f(
  \tilde{\mathcal{R}})\ ,
  \qquad {\rm where} \qquad
  f(x)=2x
  \big(1-\ln(-x/\Lambda_2)\big)
  \ .
\end{equation}

This raises the question of which is the ``physical" gravitational metric, either $g_{\mu\nu}$ (usually called the Jordan frame) or $\tilde{g}_{\mu \nu}$ (the Einstein frame).\footnote{See Refs. \cite{Faraoni:1999hp,Postma:2014vaa} for a discussion around this issue.} The distinction between the frames is important as the solutions obtained in either case are very different. For instance, if we consider a constant curvature solution for $\tilde{g}_{\mu \nu}$, the equation of motion from (\ref{eq:50}) is given by
\beq \tilde{\mathcal{R}}f'(\tilde{\mathcal{R}})-
  f(\tilde{\mathcal{R}})=0
  \qquad \Longrightarrow \qquad
  \tilde{\mathcal{R}}=0\ .\eeq
From (\ref{eq:14}), this implies the metric $\tilde{g}_{\mu \nu}$ vanishes, so that the theory does not admit a pure AdS$_2$ solution and we cannot consider the extended first law in (\ref{eq:49}). 

On the other hand, working in the Jordan frame with the metric $g_{\mu \nu}$ the action (\ref{eq:47}) allows a pure AdS$_2$ solution \cite{Frassino:2015oca}. This means it is sensible to consider the extended first law for the metric $g_{\mu \nu}$, although the derivation leading to (\ref{eq:49}) does not apply. An extended first law of black hole thermodynamics (which studies the behavior of the black hole entropy under variations of the cosmological constant) was derived in Ref. \cite{Frassino:2015oca} for the Einstein-dilaton theory in (\ref{eq:47}). In order to obtain a sensible result, the authors of Ref. \cite{Frassino:2015oca} use an unconventional approach that involves rescaling Newton's constant according to $G_{d+1}=\frac{(1-d)}{2}G_2$. Starting from the results in Ref. \cite{Kastor:2014dra}, this procedure can also be applied to derive an extended first law for perturbations of Killing horizons in the AdS$_2$ metric $g_{\mu \nu}$.

%%%%%%%%%%%%%%%%%%%%%%%%%%%%%%%%%%%%%%%
\subsection{Jackiw-Teitelboim gravity}
\label{sub:JT}
%%%%%%%%%%%%%%%%%%%%%%%%%%%%%%%%%%%%%%

In this subsection we consider the extended first law in the context of Jackiw-Teitelboim gravity \cite{Teitelboim:1983ux,Jackiw:1984je}, that correspond to an Einstein-dilaton theory that cannot be written as a purely gravitational theory of the type $\mathcal{L}=f(\mathcal{R})$. The action defining the theory can be written as
\begin{equation}\label{eq:190}
I_{JT}=I_\phi[g_{\mu \nu};\phi_0,L]=
\int d^2x\sqrt{-g}
\left[
\phi_0\mathcal{R}+
\phi(x)(\mathcal{R}+2/L^2)
\right]\ .
\end{equation}
The dilaton field $\phi(x)$ is dimensionless and there are two coupling constants that define the theory $\lambda_i=(\phi_0,L)$. As usual, the action must be supplemented with appropriate boundary terms to yield a well defined variational problem. The equations of motion can be easily computed and written as
\begin{equation}\label{eq:193}
\begin{aligned}
\mathcal{R}+2/L^2&=0\\
\left[
\nabla_\mu \nabla_\nu-\frac{g_{\mu \nu}}{L^2}
\right]\phi(x) &=0\ .
\end{aligned}
\end{equation}
The first equation fixes the Ricci scalar to a negative constant value and since the theory is two dimensional, it completely determines the Riemann tensor (\ref{eq:14}). This means the \textit{only} metric solution in JT gravity is pure ${\rm AdS}_2$. The analysis of the extended first law in JT gravity is extremely simple given that all we have to do is analyze the thermodynamic behavior of Killing horizons in pure ${\rm AdS}_2$. The theory does not admit any real black hole solution.\footnote{While the classical theory is almost trivial, interesting dynamics arise by introducing a fluctuating boundary. These boundary effects give one loop contributions to the Euclidean partition function \cite{Maldacena:2016upp,Harlow:2018tqv} and therefore lie beyond the semi-classical analysis captured by horizon thermodynamics.}

Writing the metric in global coordinates $(\sigma,\theta)$ as in (\ref{eq:13}) the only Killing horizon is generated by the vector in (\ref{eq:52}), which is time-like in the region $\theta_\pm<\theta_0\in (0,\pi)$, sketched in figure \ref{fig:2}. The equation of motion of the dilaton $\phi(x)$ can be easily solved in global coordinates and written as
\begin{equation}\label{eq:194}
\phi(\sigma,\theta)=\phi_h 
\frac{\cos(\sigma/L)\sin(\theta_0)}{\sin(\theta)}\ ,
\end{equation}
where $\phi_h> 0$ is an integration constant that gives the value of the dilaton at the horizon. The full solution is parametrized by the value of the single constant $\phi_h$.\footnote{While it seems the solution also depends on $\theta_0\in(0,\pi)$, we can use the isometries of ${\rm AdS}_2$ to fix $\theta_0=\pi/2$.}

To compute the horizon entropy we use Wald's functional (\ref{eq:45}) together with the fact that the Riemann tensor is fixed by the Ricci scalar $\mathcal{R}$ in (\ref{eq:14})
\begin{equation}\label{eq:191}
S_\xi=
4\pi
\left.\frac{\delta \mathcal{L}}{\delta R}
\right|_{\rm Horizon}=4\pi\phi_0+
4\pi \phi(x)\big|_{\theta_\pm=\theta_0}=
4\pi(\phi_0+\phi_h)\ .
\end{equation}
This agrees with the result obtained from the semi-classical computation of the Euclidean path integral \cite{Harlow:2018tqv}. The extended first law involves computing the entropy variation with respect to the coupling constants of the theory $\lambda_i=(\phi_0,L)$ and checking whether it can be written as
\begin{equation}\label{eq:195}
\delta_{\lambda_i} S_\xi=\frac{S_\xi}{a_1^\ast}
\delta_{\lambda_i} a_1^\ast\ ,
\end{equation}
where $a_1^\ast$ is some function of the coupling constants $a_1^\ast=a_1^\ast(\phi_0,L)$. In this setup we have no natural definition of $a_1^\ast$ in terms of the on-shell Lagrangian (\ref{eq:17}), so in principle we can allow any function that depends exclusively on the coupling constants $(\phi_0,L)$. However, since $a_1^\ast$ and $\phi_0$ are dimensionless quantities and $L$ has dimensions of length we have it can only depend on $\phi_0$.\footnote{Note that if we naively apply the definition of $a_1^\ast$ in (\ref{eq:17}), we get $a_1^\ast=2\phi_0$.} From the simple expression of the entropy given in (\ref{eq:191}) we can compute the entropy variation explicitly and find it is not compatible with the extended first law as written in (\ref{eq:195}) for any definition of $a_1^\ast(\phi_0)$
\begin{equation}
\delta_{\lambda_i} S_\xi=4\pi \delta \phi_0\neq \frac{S_\xi}{a_1^\ast}\delta_{\lambda_i} a_1^\ast\ .
\end{equation}
This means the form of the extended first law for JT gravity is not the same as in the previous cases we studied so far. The difference is that the solution in JT gravity depends on the additional parameter $\phi_h$, that appears in the horizon entropy and is not related to the AdS radius $L$. In the previous derivations in section \ref{sec:2} we used the fact that the pure AdS solution only depends on the radius $L$. 

We expect a similar situation for other Einstein-dilaton theories that cannot be written as pure gravity theories. For any particular theory one can still compute the variation of the horizon entropy on pure ${\rm AdS}_2$ as in (\ref{eq:191}), but there is no guarantee there exists a function $a_1^\ast=a_1^\ast(\lambda_i)$ such that it can be written as in the extended first law (\ref{eq:195}).

%%%%%%%%%%%%%%%%%%%%%%%%%%%%%%%%%%%%%
\section{Beyond pure AdS in three dimensional gravity}
\label{sec:3Dlaw}
%%%%%%%%%%%%%%%%%%%%%%%%%%%%%%%%%%%%%%

Given that all our calculations so far have been for Killing horizons in pure AdS, a natural question is whether these results can be extended to horizons in more general space-times. In this section we investigate this in the context of three dimensional gravity, making contact with some concepts in extended black hole thermodynamics \cite{Kubiznak:2016qmn}.

Consider a general three dimensional metric $g_{\mu\nu}$ which solves the equations of motion obtained from (\ref{genaction}) and admits a time-like Killing horizon generated by the vector $\xi^\mu$. The horizon entropy is obtained from Wald's functional (\ref{Waldent}) evaluated on $g_{\mu \nu}$, which for a general metric we cannot evaluate explicitly. However, three dimensional gravity theories admit interesting black hole solutions which are locally but not globally AdS, \textit{i.e.}, which satisfy
\begin{equation}\label{eq:53}
\mathcal{R}_{\mu \nu\rho \sigma}=
  -\frac{1}{L^2}
  \left(
  g_{\mu \rho}g_{\nu \sigma}-
  g_{\mu \sigma}g_{\nu \rho}
  \right)\ .
\end{equation}
For this class of black holes we can evaluate the integrand in Wald's functional using (\ref{eq:24}) and find
\begin{equation}\label{eq:38}
S_{\xi}
  \left[
  g_{\mu \nu},\lambda_i
  \right]=
  2a_2^*(\lambda_i)
  \widetilde{\mathcal{A}}
  \ ,
\end{equation}
where $\widetilde{\mathcal{A}}=\mathcal{A}_{\rm horizon}/L^{d-1}$ and $a_2^*$ in (\ref{eq:9}) is proportional to the Virasoro central charge $c$ of the dual $\text{CFT}_{2}$. This expression is equivalent to the pure AdS relation (\ref{eq:2}) evaluated at~$d=2$. 

Let us now consider the behavior of the entropy under deformations of the theory, \textit{i.e.},~${\lambda_i\rightarrow \lambda_i+\delta \lambda_i}$ in (\ref{genaction}). In this case, apart from the obvious contribution given by the coefficient $a_2^*(\lambda_i)$ in (\ref{eq:38}), we must take into account the variation of the dimensionless horizon area $\widetilde{\mathcal{A}}$. For the pure AdS metric, $\widetilde{\mathcal{A}}$ is independent of $\lambda_i$ since the metric $g_{\mu \nu}^{\rm AdS}(L)$ only depends on the dimensionful parameter $L$, so that dimensional analysis implies $\mathcal{A}_{\rm horizon}\propto L^{d-1}$. This is no longer true for more general metrics which satisfy (\ref{eq:53}) but are not globally pure AdS, as the metric can also depend on some integration constants $\left\lbrace c_j \right\rbrace$ (\textit{e.g.} mass, angular momentum, charge, etc.) so that the horizon area $\mathcal{A}_{\rm horizon}$ is no longer proportional to $L^{d-1}$. Altogether, the variation of (\ref{eq:38}) is now given by
\begin{equation}\label{eq:34}
\delta S_\xi=
  S_\xi\,\delta\left[
  \ln(a_2^*)+\ln(\widetilde{\mathcal{A}})
  \right]\ .
\end{equation}
As we will shortly see in a simple example, computing this extra variation for a particular solution is straightforward. However, while the first term involving $a_2^*$ has a clear meaning in the boundary CFT (given in (\ref{eq:32})), this is not the case for $\widetilde{\mathcal{A}}$. Only by restricting ourselves to black holes in which $\delta \widetilde{\mathcal{A}}=0$, the boundary CFT satisfies the extended first law given by
\begin{equation}\label{eq:39}
\delta \widetilde{\mathcal{A}}=0
  \qquad \Longrightarrow \qquad
  \delta S(\rho_\beta)=\beta\, \delta
  \langle H \rangle+
  \frac{S(\rho_\beta)}{a_2^*}\delta a_2^*\ ,
\end{equation}
where $\rho_\beta$ is a thermal state and we have included the usual energy term $(2\pi/\kappa)\delta E'_\xi$ in (\ref{eq:34}) which maps to $H$, the hamiltonian of the CFT. Additional conserved quantities such as angular momentum or charges, can be added to this relation in the usual way. The first law in~(\ref{eq:39}) is similar to the one obtained for the thermal state at temperature $\beta=2\pi R$ in the background~${\mathbb{R}\times \mathbb{H}^{d-1}}$ (\ref{eq:55}), with the crucial difference that $\beta$ in this case is unconstrained.  

Let us illustrate how everything works by considering a simple example in Einstein gravity
\begin{equation}\label{eq:43}
I[g_{\mu \nu};G,L]=
  \frac{1}{16\pi G}
  \int d^3x\,\sqrt{-g}\left(
  \mathcal{R}+\frac{2}{L^2}
  \right)\ .
\end{equation}
The coupling constants of the theory are $\left\lbrace \lambda_i\right\rbrace=\left\lbrace G,L \right\rbrace$, where $L$ is also the radius of the pure AdS solution. The rotating BTZ black hole solution satisfies (\ref{eq:53}) and is given by \cite{Banados:1992wn}
\begin{equation}\label{eq:44}
ds^2=-f(r)dt^2+\frac{dr^2}{f(r)}+
  r^2\Big(
  d\theta-\frac{GJ}{2r^2}dt
  \Big)^2\ ,
\end{equation}
where $f(r)=-8G M+(r/L)^2+(JG/2r)^2$. Different black holes are labeled by the integration constants ${\left\lbrace c_j \right\rbrace=\left\lbrace M,J \right\rbrace}$, which also give the global charges associated to the Killing vectors~$\partial_t$ and $\partial_\theta$ respectively.

The outer horizon radius $r_+$ is obtained from $f(r_+)=0$ and is a non-trivial function of~$(G,L,M,J)$. We can easily write the dimensionless horizon area $\widetilde{\mathcal{A}}$ in terms of $r_+$
\begin{equation}\label{eq:35}
\widetilde{\mathcal{A}}=
  \frac{2\pi r_+}{L}=
  4\pi \sqrt{MG}
  \left[
  1+\sqrt{1-\left(
  \frac{J}{8ML}
  \right)^2}\,
  \right]^{1/2}.
\end{equation}
This expression depends explicitly on both $G$ and $L$, meaning that the second term in (\ref{eq:34}) gives a non-trivial contribution, which we can easily write explicitly. However, if we consider the static black hole $J=0$ we get $\widetilde{\mathcal{A}}=4\pi\sqrt{2MG}$, which is independent of $L$. Therefore, if we restrict to variations of $L$ (while keeping $G$ fixed), we obtain the extended first law given in (\ref{eq:39}).

%%%%%%%%%%%%%%%%%%%%%%%%%%%%%%%%%%%%
\subsection{Extended thermodynamics and volume}
%%%%%%%%%%%%%%%%%%%%%%%%%%%%%%%%%%%%
 
Let us now restrict to a particular type of theory deformation, in which we take the radius of the pure AdS solution $L$ as one of the coupling constants defining the theory and consider~$\delta(\lambda_i,L)=(0,\delta L)$. This corresponds to the variations studied in the extended black hole thermodynamics \cite{Kubiznak:2016qmn}, in which the thermodynamic pressure is identified with $L$ according to~${p\equiv d(d-1)/(16\pi G L^2)}$. Its conjugate variable is referred as the volume $V$ and can be defined from the entropy as
\begin{equation}\label{eq:36}
V\equiv -T \frac{\partial S_\xi}{\partial p}=
  -T S_\xi
  \frac{\partial }{\partial p}
  \left[
  \ln(a_2^*)+\ln(\widetilde{\mathcal{A}})
  \right]\ .
\end{equation}
where the second equality is obtained from (\ref{eq:34}). The $p$ derivative is computed while keeping all the remaining parameters fixed. 

This volume formula holds for locally AdS black holes in any three dimensional theory of gravity. Similar to (\ref{eq:34}), there are two distinct contributions to the volume. While the variation of $a_2^*$ has a natural boundary interpretation in terms of the number of degrees of freedom, the dimensionless area $\widetilde{\mathcal{A}}$ does not. For cases in which $\widetilde{\mathcal{A}}$ is independent of $L$, the thermodynamic volume takes the following simple form
\begin{equation}\label{eq:37}
\frac{\partial \widetilde{\mathcal{A}}}
{\partial L}=0
\quad \Longrightarrow \quad
V=-\left(\frac{T S_\xi}{a_2^*}\right)
  \frac{\partial a_2^*}{\partial p}\ .
\end{equation}
This gives a class of three dimensional black holes whose thermodynamic volume is directly related to changing the central charge of the boundary CFT. Since the meaning of $V$ for the boundary theory is not completely understood (see Refs. \cite{Dolan:2013dga,Johnson:2014yja,Dolan:2014cja,Kastor:2014dra,Caceres:2016xjz,Couch:2016exn,Johnson:2019wcq}), this formula might help give further insights. Let us use it in some concrete examples to compute the volume of some black hole solutions.

%%%%%%%%%%%%%%%%%%%%%%%%%%%%%%%%%%%%%%
\subsubsection*{Thermodynamic volume in Einstein gravity}
%%%%%%%%%%%%%%%%%%%%%%%%%%%%%%%%%%%%%%

Consider the simple setup of a BTZ black hole (\ref{eq:44}) in Einstein gravity (\ref{eq:43}). As previously noted, for the static black hole $J=0$ the dimensionless horizon area $\widetilde{\mathcal{A}}$ in (\ref{eq:35}) is independent of $L$, meaning that we can directly use the volume formula in (\ref{eq:37}). Simple calculations give~$a_2^*=L/8G$ and $T=r_+/2\pi L^2$, so that we can compute the volume as
\begin{equation}
V_{J=0}=-\left(\frac{T S_\xi}{a_2^*}\right)
  \frac{\partial a_2^*}{\partial p}=
   \pi r_+^2\ .
\end{equation}
which agrees with the result obtained from a more standard approach in extended thermodynamics \cite{Frassino:2015oca}. 

For the rotating BTZ solution with $J\neq 0$ the dimensionless horizon area $\widetilde{\mathcal{A}}$ in (\ref{eq:35}) is a non-trivial function of $L$, meaning that we must use the more general volume formula in~(\ref{eq:36}). Although the calculation in this case is slightly more involved, the final result is again very simple and given by
\begin{equation}
V_{J\neq 0}=
  -T S_\xi
  \frac{\partial }{\partial p}
  \left[
  \ln(a_2^*)+\ln(\widetilde{\mathcal{A}})
  \right]=
  \pi r_+^2\ ,
\end{equation}
in agreement with the previously known relation \cite{Frassino:2015oca}. It is interesting to see that the extra variation with respect to $\widetilde{\mathcal{A}}$ is exactly what is needed in order to obtain this simple final answer. An interesting microscopic analysis of this expression was recently given in Ref.~\cite{Johnson:2019wcq}.\footnote{We should mention that while the charged BTZ black hole in Einstein-Maxwell theory \cite{Martinez:1999qi} is not locally AdS (\ref{eq:53}), if we naively apply the volume formula in (\ref{eq:37}) we obtain ${V=\pi r_+^2-\pi(QL/2)^2}$, which agrees with the previously known result \cite{Frassino:2015oca}. The reason it works is due to the fact that in Einstein gravity Wald's entropy functional always reduces to the Bekenstein-Hawking area expression, \textit{i.e.} $S_\xi=\mathcal{A}/4G$. For higher curvature theories we do not expect the volume formula (\ref{eq:36}) to reproduce the correct result for the charged black hole.}

%%%%%%%%%%%%%%%%%%%%%%%%%%%%%%%%%%%%%%
\subsubsection*{Thermodynamic volume in higher curvature theories}
%%%%%%%%%%%%%%%%%%%%%%%%%%%%%%%%%%%%%%

Since the volume formula (\ref{eq:36}) is particularly powerful in the context of higher curvature gravity theories, let us apply it in an example by considering the following generalization of new massive gravity \cite{Bergshoeff:2009hq,Bergshoeff:2009aq,Sinha:2010ai}
\beq I[g_{\mu \nu}]=\frac{1}{16\pi G}\int d^{3}x\sqrt{-g}\left(\mathcal{R}+\frac{2}{\ell^{2}}+\ell^{2}\mathcal{R}_{2}+\ell^{4}\mathcal{R}_{3}\right)\;,\label{NMGextaction}\eeq
where 
\beq
\begin{split}
& \mathcal{R}_{2}=4(\lambda_{1}\mathcal{R}_{\mu \nu}\mathcal{R}^{\mu \nu}+\lambda_{2}\mathcal{R}^{2})\;,\\
& \mathcal{R}_{3}=\frac{17}{12}
(\mu_{1}\mathcal{R}^{\nu}_{\;\mu}\mathcal{R}^{\rho}_{\;\nu}\mathcal{R}^{\mu}_{\;\rho}
+\mu_{2}\mathcal{R}_{\mu \nu}\mathcal{R}^{\mu \nu}\mathcal{R}
+\mu_{3} \mathcal{R}^{3})\;.\label{curves}
\end{split}
\eeq
The coupling constants of the theory are given by $\left\lbrace G,\ell,\lambda_1,\lambda_2,\mu_i \right\rbrace$ with $i=1,2,3$, where new massive gravity \cite{Bergshoeff:2009hq,Bergshoeff:2009aq} is obtained by setting $\mu_i=0$ and $\lambda_{2}=-3\lambda_{1}/8$. 

To apply the volume formula in (\ref{eq:36}) we must first compute the $a_2^*$ factor, which depends on the pure AdS solution of the theory. We can find such solution by varying the action~(\ref{NMGextaction}) with respect to the metric, which gives the following equations of motion \cite{Sinha:2010ai}
\begin{equation}\label{eqnsofmotNMG}
\mathcal{R}_{\mu \nu}
-\frac{1}{2}\mathcal{R}g_{\mu \nu}-
\frac{1}{\ell^{2}}g_{\mu \nu}-
H_{\mu \nu}=0\;,
\end{equation}
where
\begin{equation}
\begin{split}
H_{\mu \nu}&=4\ell^{2}
\biggr[\lambda_{1}
\left(-2\mathcal{R}^{\rho}_{\;\mu}\mathcal{R}_{\rho \nu}
+\frac{1}{2}g_{\mu \nu}
\mathcal{R}_{\rho \sigma}
\mathcal{R}^{\rho \sigma}\right)
+
\lambda_{2}\left(-2\mathcal{R}\mathcal{R}_{\mu \nu}+\frac{1}{2}g_{\mu \nu}\mathcal{R}^{2}\right)
\biggr]\\
&+\frac{17}{12}\ell^{4}
\biggr[
\mu_{1}
\left(-3\mathcal{R}_{\mu \rho}\mathcal{R}^{\rho}_{\;\sigma}\mathcal{R}^{\sigma}_{\;\nu}
+\frac{1}{2}g_{\mu \nu}\mathcal{R}^{\rho}_{\;\sigma}\mathcal{R}^{\alpha}_{\;\rho}\mathcal{R}^{\sigma}_{\;\alpha}\right)
+\mu_{3}
\left(-3\mathcal{R}^{2}\mathcal{R}_{\mu \nu}+\frac{1}{2}g_{\mu \nu}\mathcal{R}^{3}\right)
\\
&+
\mu_{2}
\left(-\mathcal{R}^{\rho}_{\;\sigma}\mathcal{R}^{\sigma}_{\;\rho }\mathcal{R}_{\mu \nu}-2\mathcal{R}\mathcal{R}_{\mu \rho}\mathcal{R}^{\rho}_{\;\nu }+\frac{1}{2}g_{\mu \nu}\mathcal{R}
\mathcal{R}_{\rho \sigma}
\mathcal{R}^{\rho \sigma}\right)
\biggr]
+\mathcal{O}(\nabla^{2}\mathcal{R},\nabla^{2}\mathcal{R}^{2},...)\; ,
\end{split}
\end{equation}
and we are omitting derivative terms that do not contribute to the pure AdS solution. 

We can evaluate these complicated terms in a pure AdS metric $g_{\mu \nu}^{\rm AdS}(L)$ of some radius~$L$ using that it is a maximally symmetric space-time (\ref{eq:53}). Taking the trace of (\ref{eqnsofmotNMG}) and writing the AdS radius as $L=\ell/\sqrt{f_\infty}$ we obtain the following algebraic constraint for the factor $f_\infty$
\beq 
L=\ell/\sqrt{f_\infty}
\qquad \Longrightarrow \qquad
1-f_{\infty}-8f_{\infty}^{2}(\lambda_{1}+3\lambda_{2})+17f_{\infty}^{3}(\mu_{1}+3\mu_{2}+9\mu_{3})=0\;.\label{polyfinf}\eeq
The solution $f_\infty$ of this algebraic equation that is smoothly connected to Einstein gravity determines the pure AdS radius $L$. We can now write $a_2^*$ from (\ref{eq:9}) by evaluating the Lagrangian density (\ref{NMGextaction}) in AdS, so that we find
\begin{equation}\label{eq:54}
a_2^*=
 -
 \frac{1}{2}\pi L^3
 \mathcal{L}\big|_{\rm AdS}=
\frac{L}{8 G}\left[1-16f_{\infty}(\lambda_{1}+3\lambda_{2})+17f_{\infty}^{2}(\mu_{1}+3\mu_{2}+9\mu_{3})\right]\;,
\end{equation}
where we have used
\begin{equation}
\mathcal{R}_{2}=
\frac{48}{L^{4}}
(\lambda_{1}+3\lambda_{2})\ ,
\qquad \qquad
\mathcal{R}_{3}=
-\frac{34}{L^{6}}\left(\mu_{1}+3\mu_{2}+9\mu_{3}\right)\ .
\end{equation}
The expression for $a_2^\ast$ and the constraint in (\ref{polyfinf}) defining $f_\infty$ reduce to the ones given in Ref. \cite{Sinha:2010ai} when setting~$\lambda_2=-3\lambda_1/8$ and~${(\mu_1,\mu_2)=\mu_3(64,-72)/17}$. Moreover, if we take~${\lambda_1=\lambda_2=\mu_i=0}$ we get $f_\infty=1$ and $a_2^\ast=L/8G$, in agreement with the Einstein gravity results. Notice that the dependence of $a_2^\ast$ with the AdS radius $L$ is linear, as in the Einstein case.

We can now consider a black hole solution for this theory. Given that the BTZ black hole in (\ref{eq:44}) is locally AdS, it solves the equations of motion in (\ref{eqnsofmotNMG}) as long as we take~$L$ according to (\ref{polyfinf}). The horizon entropy is obtained from (\ref{eq:38}) with $a_2^*$ and $\widetilde{\mathcal{A}}$ as given in~(\ref{eq:54}) and (\ref{eq:35}). For the rotating solution with $J\neq 0$ we can use the volume formula in~(\ref{eq:36}) and find
\begin{equation}\label{eq:56}
V_{J\neq 0}=
  \pi r_{+}^{2}
  \left[1-16f_{\infty}(\lambda_{1}+3\lambda_{2})+f^{2}_{\infty}(\mu_{1}+3\mu_{2}+9\mu_{3})\right]\;.
\end{equation}
To our knowledge, higher curvature contributions to the BTZ thermodynamic volume have not been computed before.

%%%%%%%%%%%%%%%%%%%%%%%%%%%%%%%%%%%%%%%%%%%
\section{Discussion}
\label{sec:4}
%%%%%%%%%%%%%%%%%%%%%%%%%%%%%%%%%%%%%%%%%%%

The extended first law of entanglement has been previosuly derived for the Minkowski vacuum reduced to a ball by considering particular gravity theories in the bulk \cite{Kastor:2014dra,Kastor:2016bph,Caceres:2016xjz,Lan:2017xcl}. In this work, we have shown a novel and simple procedure that generalizes the proof to arbitrary gravity theories in the bulk and new setups in the boundary CFT. From the bulk perspective we have found no obstructions in working in two dimensional gravity and also obtain some intriguing results concerning extended black hole thermodynamics in three dimensions. Let us discuss some additional aspects regarding the calculations in the main text.

%%%%%%%%%%%%%%%%%%%%%%%%%%%%%%%%%%%%%%%%
\subsection*{Divergent terms in the extended first law of entanglement}
%%%%%%%%%%%%%%%%%%%%%%%%%%%%%%%%%%%%%%%%

One important feature of the ordinary first law of entanglement $\delta S_{EE}=\delta \langle K_B \rangle$ is that although the entanglement entropy always diverges, the left-hand side is well defined since the difference between entropies associated to different states is finite.\footnote{As shown in Ref. \cite{Marolf:2016dob} this is not entirely true, since there are cases in which the entanglement entropy acquires state dependent divergences, so that $\delta S_{EE}$ diverges. However, the relative entropy remains finite.} For the extended first law of entanglement this is no longer the case. Let us consider a variation of the theory without perturbing the state, so that the first term on the right-hand side of (\ref{eq:12}) drops out and we are left with
\begin{equation}\label{eq:41}
S_{EE}(\bar{\rho})-S_{EE}(\rho)=
  \frac{S_{EE}(\rho)}{a_d^\ast}
  \delta a_d^\ast\ .
\end{equation}
Both sides of this equality diverge, the left-hand side due to the fact that the divergences of the entanglement entropies corresponding to different theories do not cancel each other. This raises the question regarding how we should interpret (\ref{eq:41}), which seems to depend on the regularization procedure.

Let us illustrate the issue by considering the simple case of the Minkowski vacuum reduced to a ball of radius $R$ in $d=3$, where the entanglement entropy is \cite{Casini:2011kv}
\begin{equation}\label{eq:42}
S_{EE}(\rho_B)=
  \mu_1\frac{R}{\epsilon}-2 \pi a_3^*\ ,
\end{equation}
with $\mu_1$ a dimensionless and non-universal constant and $a_3^*$ given by (\ref{eq:32}). The short distance cut-off $\epsilon$ can be properly defined using mutual information, see Ref. \cite{Casini:2015woa}. If we consider the same setup but for a CFT in which ${\bar{a}_3^*=a_3^*-\delta a_3^*}$, the entanglement entropy is given by
\beq S_{EE}(\bar{\rho}_B)=
  \bar{\mu}_1\frac{R}{\bar{\epsilon}}-2 \pi \bar{a}_3^*\ ,\eeq
where the cut-off $\bar{\epsilon}$ and the constant $\bar{\mu}_1$ are not necessarily related to the ones appearing in~(\ref{eq:42}). 

How should we understand (\ref{eq:41}) in this context? A practical approach is to simply ignore the non-universal contributions to the entanglement entropy and regard (\ref{eq:41}) as a relation between the universal terms, where it is clear that the extended first law is satisfied. A different procedure is instead given by relating the cut-offs of each theory in a particular way such that the extended first law is satisfied to every order. Assuming there is a relation~$\epsilon=\epsilon(\bar{\epsilon})$ which can be expanded around the origin as
\beq \epsilon(\bar{\epsilon})=
  \bar{\epsilon}\left(
  b_0+b_2(\bar{\epsilon}/R)^2+
  b_4(\bar{\epsilon}/R)^4+\dots
  \right)\ ,\eeq
we can fix the coefficients $b_{2n}$ such that (\ref{eq:41}) is satisfied to every order. For the case of a ball in three dimensional Minkowski we find
\beq \epsilon(\bar{\epsilon})=
  \bar{\epsilon}\,
  \frac{\mu_1}{\bar{\mu}_1}
  \left(1-\delta \ln(a_3^*)\right)+\dots\ ,\eeq
where higher order terms are unconstrained. An analogous construction can be considered for the higher dimensional case and other setups in the CFT. This subtle aspect regarding the extended first law of entanglement has not been previously discussed in the literature.

%%%%%%%%%%%%%%%%%%%%%%%%%%%%%%%%%
\subsection*{Extended first law of entanglement for general setups}
%%%%%%%%%%%%%%%%%%%%%%%%%%%%%%%%%%%%%%

Given that we have shown that the extended first law of entanglement holds in a wide variety of setups, a natural question is whether it is valid for arbitrary CFTs, regions and states. While the ordinary first law follows from positivity of relative entropy \cite{Blanco:2013joa} and therefore holds in full generality, the extended version can only be formulated for CFTs since the coefficient $a_d^\ast$ in even dimensions is only defined for conformal theories (\ref{eq:32}). Although trying to directly prove the extended first law for arbitrary CFTs seems a complicated task, we can check whether the results for the entanglement entropy present in the literature are consistent with (\ref{eq:ext}), which essentially implies $S_{EE}\propto a_d^\ast$ to first order in $a_d^\ast$.

Let us consider two dimensional CFTs, where $a_2^\ast$ is proportional to the Virasoro central charge $c$. For the vacuum entropy associated to any number of disjoint intervals of a holographic CFT in Minkowski space, Refs. \cite{Ryu:2006ef,Hartman:2013mia,Faulkner:2013yia} showed that $S_{EE}\propto a_2^\ast$. The same is true for a thermal state reduced to an interval \cite{Calabrese:2004eu} and analogous setups in curved backgrounds \cite{Cardy:2016fqc}. For more general situations, the entanglement entropy is only known for particular CFTs, mostly free theories. In each of these cases the entropy depends on the details of the theory  in a complicated way. However, we are not aware of any result where the entanglement entropy in two dimensions is not proportional to the central charge and, consequently, in contradiction with (\ref{eq:ext}).

For space-time dimensions larger than two, it becomes evident that the extended first law of entanglement as written in (\ref{eq:ext}) cannot hold in full generality. The simplest example is to consider the Minkowski vacuum in $d=4$ reduced to a cylinder. Here the entanglement entropy is independent of $a_d^\ast$ and is instead proportional to the coefficient appearing in the square of the Weyl tensor in the trace anomaly \cite{Solodukhin:2008dh}. For more complicated regions the entropy is a combination of these coefficients. While this shows the extended first law as written in (\ref{eq:ext}) cannot hold in general for $d=4$, it suggests the following generalization might still be true\footnote{We thank Manus Visser for suggesting this generalization.}
\begin{equation}
\delta S_{EE}=\delta\langle K_B \rangle+
\sum_{i}\frac{S_{EE}}{a_i}\delta a_i\ ,
\end{equation}
where $B$ is a region in four-dimensional Minkowski and $a_i$ are the coefficients of the terms appearing in the trace anomaly (see for example \cite{Myers:2010tj}). This generalization has a better chance of applying to more general regions.

It would be interesting to understand how holography is able to capture the extended first law of entanglement in these more general cases where it is expected to hold. The $d=2$ case stands out as the simplest one in which concrete progress might be possible, maybe using similar techniques as the ones developed in \cite{Faulkner:2013yia}. This deserves further study, in order to determine whether a general derivation of the extended first law of entanglement in this context is possible.

%%%%%%%%%%%%%%%%%%%%%%%%%%%%%%%%%%%%
\subsection*{Bulk constraints from extended first law of entanglement}
%%%%%%%%%%%%%%%%%%%%%%%%%%%%%%%%%%%%

Assuming the RT holographic formula for entanglement entropy together with the ordinary first law of entanglement in the boundary, implies Einstein's bulk equations about a perturbed AdS background. What are the consequences of assuming the extended first law of entanglement instead?\footnote{We thank Manus Visser for suggesting this question} 

Let us address this question in the simplest setup of ${\rm AdS}_3/{\rm CFT}_2$, where the bulk theory is described by Einstein gravity, so that the coupling constants are $\lambda_i=(G,L)$. Let us assume (the non-trivial statement that) the extended first law of entanglement holds in the boundary CFT for arbitrary states $\rho$ and regions $B$, together with the RT formula
\begin{equation}\label{eq:201}
\delta S_{EE}=\delta \langle K_B \rangle+
\frac{S_{EE}}{c}\delta c\ ,
\qquad \qquad
S_{EE}=
\frac{A(\gamma_{\rm ext})}{4G} \ ,
\end{equation}
where $\gamma_{\rm ext}$ is an extremal bulk curve homologous to the region $B$ at the boundary. Using that in Einstein gravity the central charge $c$ is given by $c=3L/2G$, the ``extended" contribution of the first law of entanglement on the bulk becomes
\begin{equation}\label{eq:200}
\delta_{\lambda_i}\left(
\frac{A(\gamma_{\rm ext})}{4G}
\right)=
\frac{A(\gamma_{\rm ext})}{4G}
\delta_{\lambda_i}\ln(L/G)
\qquad \Longrightarrow \qquad
A(\gamma_{\rm ext})\propto L\ .
\end{equation}
The extended first law of entanglement translates into the statement that the length of the extremal curve on the bulk is proportional to the AdS radius $L$. 

If the boundary state is the vacuum $\ket{0}$ the bulk metric is pure ${\rm AdS}_3$, which only depends on $L$, and $A(\gamma_{\rm ext})\propto L$ immediately follows from dimensional analysis. The constraint becomes more interesting when considering excited states at the boundary, such as a thermal state $\rho(\beta)$ with inverse temperature $\beta$. In this case we can easily compute $A(\gamma_{\rm ext})$ and find the non-trivial statement $A(\gamma_{\rm ext})\propto L$ is indeed true \cite{Ryu:2006bv}. For more general setups this gives a bulk constraint coming from the boundary extended first law of entanglement.

It is also interesting to consider the inverse logic. We can directly compute $A(\gamma_{\rm ext})$ for complicated holographic setups and check whether the end result is proportional to $L$. This could help understand in which situations the extended first law of entanglement holds for the boundary theory.

%%%%%%%%%%%%%%%%%%%%%%%%%%%%%%%%%%%%
\subsection*{Extended first law in a single dimension}
%%%%%%%%%%%%%%%%%%%%%%%%%%%%%%%%%%%%

In this work we have explored the extended first law for two-dimensional gravitational theories. While we have shown interesting results can be obtained from the gravitational perspective we have not analyzed the boundary interpretation of our calculations. In future work it would be interesting to study this further, maybe in the setup of JT gravity that it has been recently understood as a holographic description of an ensemble average of SYK models \cite{Saad:2019lba}.

JT gravity also offers an arena to study the relation between quantum bulk effects and the extended first law of entanglement at the boundary. While in general it is very difficult to take these contributions into account, this simple setup allows for very explicit calculations in the bulk \cite{Jafferis:2019wkd}. Hence, it might be possible to write down an extended first law that incorporates bulk quantum corrections. On a more speculative note, it would be interesting to investigate the extended first law in dynamical space-times, in the hope it sheds a new perspective regarding recent progress on the black hole information paradox \cite{Penington:2019npb,Almheiri:2019psf}.

\subsection*{Three dimensional gravity and thermodynamic volume}
%%%%%%%%%%%%%%%%%%%%%%%%%%%%%%

For three dimensional bulk duals we have derived a modification of the extended first law~(\ref{eq:39}) that holds for space-times that are not necessarily (globally) pure AdS, such as the BTZ black hole. In the context of extended black hole thermodynamics, we obtain a curious formula for the thermodynamic volume (\ref{eq:36}), which we verified gives the correct expressions found using standard means. In particular, we obtain a result for the thermodynamic volume of the BTZ black hole in a higher curvature theory of gravity (\ref{eq:56}).

It would be interesting to see whether the formula for the thermodynamic volume in~(\ref{eq:36}) provides anything new to the field of extended thermodynamics. Particularly, it would be beneficial to see if it gives another microscopic viewpoint of $V$, along the lines of~\cite{Johnson:2019wcq}. In Ref.~\cite{Johnson:2019wcq} it was shown that the thermodynamic volume sometimes constrains the number of available CFT states dual to $\text{AdS}_{3}$ gravity, revealing that the Bekenstein-Hawking entropy (given by the Cardy formula) overcounts the number of CFT degrees of freedom. This chain of reasoning provides a microscopic explanation for black hole super-entropicity, a designation for black holes whose entropy exceeds that of Schwarzschild-AdS, and violate the reverse isoperimetric inequality  \cite{Cvetic:2010jb}. In three space-time dimensions, the reverse isoperimetric inequality takes the form
\beq \pi V\geq 4S^{2}G^{2}\;.\label{reviso3d}\eeq
When we input our expression for the volume in (\ref{eq:36}), the reverse isoperimetric inequality imposes a lower bound on the $L$ derivative of $\log(a_{2}^{\ast})$,
\beq \frac{\partial}{\partial L}\left[\log(a^{\ast}_{2})+\log(\tilde{\mathcal{A}})\right]\geq\frac{SG}{\pi^{2}L^{3}T}\geq0\;.\eeq
Black holes which satisfy this inequality, \emph{e.g.}, rotating BTZ, are said to be sub-entropic. Super-entropic black holes, such as the charged BTZ, violate the inequality (\ref{reviso3d}) and impose the following upper bound
\beq \frac{\partial}{\partial L}\left[\log(a^{\ast}_{2})+\log(\tilde{\mathcal{A}})\right]\leq\frac{SG}{\pi^{2}L^{3}T}\;.\eeq
Since $a^{\ast}_{2}$ relates to the number of degrees of freedom of the dual $\text{CFT}_{2}$, these bounds are expected to tell us something about the availability of CFT microstates to be counted by the Cardy formula. It would be interesting to study these bounds in further detail, where $\tilde{\mathcal{A}}$ might acquire a boundary interpretation.

\vspace{10pt}
\begin{acknowledgments}
It is a pleasure to thank Clifford V. Johnson and Robie Hennigar for useful comments and discussions. We are also grateful for Manus Visser, whose suggestions and questions improved the quality of this paper. FR is partially supported by DOE grant DE-SC0011687. 
\end{acknowledgments}

\bibliographystyle{JHEP}
\bibliography{sample}

\end{document}